\documentclass[aps,twocolumn,pra,showpacs,tightenlines,superscriptaddress,]{revtex4-1}
\usepackage{amsmath}
\usepackage{graphicx}
\usepackage{color}
\usepackage{amsfonts}
\usepackage{epsfig}
\usepackage{txfonts}
\usepackage{bm}
\usepackage[colorlinks,citecolor=blue]{hyperref}
\begin{document}

\title{Simultaneous ground-state cooling of two levitated nanoparticles by coherent scattering}
\author{Yi Xu}
\affiliation{Key Laboratory of Low-Dimensional Quantum Structures and Quantum Control of Ministry of Education, Key Laboratory for Matter Microstructure and Function of Hunan Province, Department of Physics and Synergetic Innovation Center for Quantum Effects and Applications, Hunan Normal University, Changsha 410081, China}
\author{Yu-Hong Liu}
\affiliation{Key Laboratory of Low-Dimensional Quantum Structures and Quantum Control of Ministry of Education, Key Laboratory for Matter Microstructure and Function of Hunan Province, Department of Physics and Synergetic Innovation Center for Quantum Effects and Applications, Hunan Normal University, Changsha 410081, China}
\author{Cheng Liu}
\affiliation{Key Laboratory of Low-Dimensional Quantum Structures and Quantum Control of Ministry of Education, Key Laboratory for Matter Microstructure and Function of Hunan Province, Department of Physics and Synergetic Innovation Center for Quantum Effects and Applications, Hunan Normal University, Changsha 410081, China}
\author{Jie-Qiao Liao}
\email{Corresponding author: jqliao@hunnu.edu.cn}
\affiliation{Key Laboratory of Low-Dimensional Quantum Structures and Quantum Control of Ministry of Education, Key Laboratory for Matter Microstructure and Function of Hunan Province, Department of Physics and Synergetic Innovation Center for Quantum Effects and Applications, Hunan Normal University, Changsha 410081, China}
\affiliation{Institute of Interdisciplinary Studies, Hunan Normal University, Changsha, 410081, China}

\begin{abstract}
Simultaneous ground-state cooling of two levitated nanoparticles is a crucial prerequisite for investigation
of macroscopic quantum effects such as quantum entanglement and quantum correlation involving translational motion of particles. Here we consider a coupled
cavity-levitated-particle system and present a detailed derivation of its Hamiltonian. We find that the $y$-direction motions of the two particles are decoupled from the cavity field and both the $x$- and $z$-direction motions, and that the $z$-direction motions can be further decoupled from the cavity field and the $x$-direction motions by choosing proper locations of the particles. We study the simultaneous cooling of these mechanical modes in both the three-mode and five-mode cavity-levitated optomechanical models. It is found that there exists a dark-mode effect when the two tweezers have the same powers, which suppress the simultaneous ground-state cooling. Nevertheless, the simultaneous ground-state cooling of these modes can be realized by breaking the dark-mode effect under proper parameters. Our system provides a
versatile platform to study quantum effects and applications in cavity-levitated optomechanical systems.
\end{abstract}

\date{\today}
\maketitle

\section{Introduction}
With the development of micro- and nano-fabrication techniques,
great advances have recently been achieved in cavity
optomechanics, especially on the fundamentals of quantum physics and modern quantum
technology~\cite{Aspelmeyer2014,MAPR2014}. The optically levitated
particles, as a kind of novel optomechanical platform,
have attracted much attention from the communities of quantum optics and quantum information~\cite{CSC2021,JRPP2020,Gaxv2307}.
In 1970s, it has been discovered that the particles can be levitated
by focusing beams of light~\cite{Ask1,Ask2,Ask3}, and this discovery has
played a crucial role in advancing the field of atom trapping and cooling~\cite{ARMP2009}. In recent years,
much attention has been paid to quantum manipulation of the translation and rotation of the center-of-mass
of particles, and great advances have been made in this platform, such as the realization of a controllable torque induced by the spins of atoms embedded in a microscale object~\cite{TN2020},
the measurement of the Brownian motion of micrometer-sized beads~\cite{TS2010}, and the cooling of the motion of particles into the quantum ground state~\cite{US2020,LN2021,FN2021}. The levitated particles can also be utilized for quantum precision measurements, including acceleration measurement~\cite{FPRA2017,APRA2018}%, force sensing measurement~\cite{TS2010}
, mass measurement~\cite{YPRL2020}, and gyroscope~\cite{RPRL2018,JPRL2018}.

Levitated nanoparticles were conceived as a candidate to explore macroscopic
quantum phenomena~\cite{PNAS2010,NJP2010,TLNP2011,APRL2010,JPRL2012,NPNAS2013,JNN2014,DQST2021,TNP2023}. This is because the nanoparticles are considered as a kind of macroscopic quantum system, and they can be levitated in a high vacuum~\cite{NJP2010,PNAS2010,TLNP2011,Uaxv1902}, which reduces the
thermal contact between mechanical motion and environment. As a
result, these systems have exceptionally high mechanical quality factors, and are considered as an excellent candidate for studying low-dissipation optomechanics.
The first step to exploring quantum effects in macroscopic mechanical systems
is the cooling of the mechanical systems to their ground states~\cite{IPRL2007,FPRL2007,JN2011,CPRL2019,LiuPRA2022}.
It has been reported that the levitated particles can be significantly cooled via feedback cooling~\cite{TLNP2011,JPRL2012,APRL2010,RPRA2018,GPRL2019,FPRL2019}
and sideband cooling~\cite{PRA2010,PRA2011,PNC2013,JPRL2015,PPRL2016,APRR2022}.
The standard sideband-cooling method in optomechanical systems
typically requires an externally red-detuned pumping field to remove the energy from the particles~\cite{NPNAS2013,NPRL1902,Uaxv1902}. However, high driving
powers will lead to the trapping of cavity fields for the optically levitated systems, and then reducing the cooling rate~\cite{Uaxv1902}.
In addition, the laser-phase noise can hinder ground-state cooling at the relevant frequencies of the
trapped nanoparticles~\cite{PPRA2009,ANJP2012,ANJP2013}. To overcome these challenges,
the coherent scattering technique has been introduced into the levitated particle
systems~\cite{UPRL2019,DPRL2019,CPRA2019,US2020,JNP2023}, drawing from atomic physics experiments~\cite{VPRL2000}. This method harnesses higher optical trapping powers and larger
particles to achieve stronger coupling strengths~\cite{NC2021}, thereby paving the way to
ultra strong coupling~\cite{Kaxv2305} and leading to novel quantum optomechanical
effects. These works were mostly based on the cooling of a single particle.
In parallel, an increasing
amount of research is focusing on the field of multi-levitated particles~\cite{CPRL2017,YO2018,HPRA2020,ANJP2020,VO2021,YO2022,HPRL2022,TPRR2023,Harx2023,JVarx2023,JNN2023,Varx2023,Marx2023,CarX2023}.
Compared with other mechanical oscillator arrays, the array of optically levitated particles has better controllability~\cite{SA2020,ANPJ2022,JPR2023,CSC2023,YSC2023}.
Motivated by these advances, we are committed to study the simultaneous ground-state cooling of multiple levitated particles and to explore more novel quantum effects.

In this paper, we study the simultaneous ground-state cooling of two levitated particles coupled to a cavity field.
Similar to a single-particle case, the photon enters the cavity via the scattering
process, which provides the mechanism for cooling the center-of-mass motions of the particles. For multiple particles levitated simultaneously,
the mechanical effect of the scattered light between the particles has been ignored
in the past. However, the recent experimental observations indicate that this scattering effect cannot be neglected for some cases~\cite{MPRL1989,VARB2006,KRMP2010,JS2022}. The redistributed light field greatly affects the
equilibrium position of the particles. Therefore, we consider the optical binding effect
between the two nanoparticles. In particular, we find that the $y$-direction motions of the two particles decouple from other degrees of freedom in this system. We also find that, when the two nanoparticles are located at the cavity nodes,
the cavity mode only couples to the $x$ modes of the particles.
Benefiting from the extremal isolation of the system, the simultaneous ground-state cooling of the center-of-mass motions of the
two nanoparticles along $x$-axis can be realized.
When the two nanoparticles are not located at the specific positions, both the $x$ mode
and the $z$ mode are coupled to the cavity mode, then the two modes of the two nanoparticles
can also be cooled into their ground states. In addition, we find that there exists the dark-mode
effect when the powers of the two tweezers are identical, and the dark-mode effect will
suppress the cooling of the system. By choosing proper parameters to avoid the dark-mode existing condition,
then the dark-mode effect can be broken, and the simultaneous ground-state cooling can be achieved.

The rest of the paper is organized as follows. In Sec.~\ref{Physical model and Hamitonian},
we introduce the system consisting of two levitated nanoparticles trapped in a Fabry-P\'{e}rot cavity, and
analytically derive the Hamiltonians. In Sec.~\ref{Cooling to two modes ground-state},
we investigate the simultaneous ground-state cooling of the $x$-direction motions of the two particles, which are located at the nodes of the cavity.
In Sec.~\ref{Transition to four modes cooling}, we study the simultaneous ground-state cooling of both the $x$- and $z$-direction motions of the two nanoparticles in a general case. Finally, we present some discussions and a brief conclusion in Sec.~\ref{conclusion}. An Appendix is presented to show the detailed derivation for a part of the interaction Hamiltonians.

\section{Physical Model and Hamiltonians}\label{Physical model and Hamitonian}
\begin{figure}[t!]
\center\includegraphics[width=0.48\textwidth]{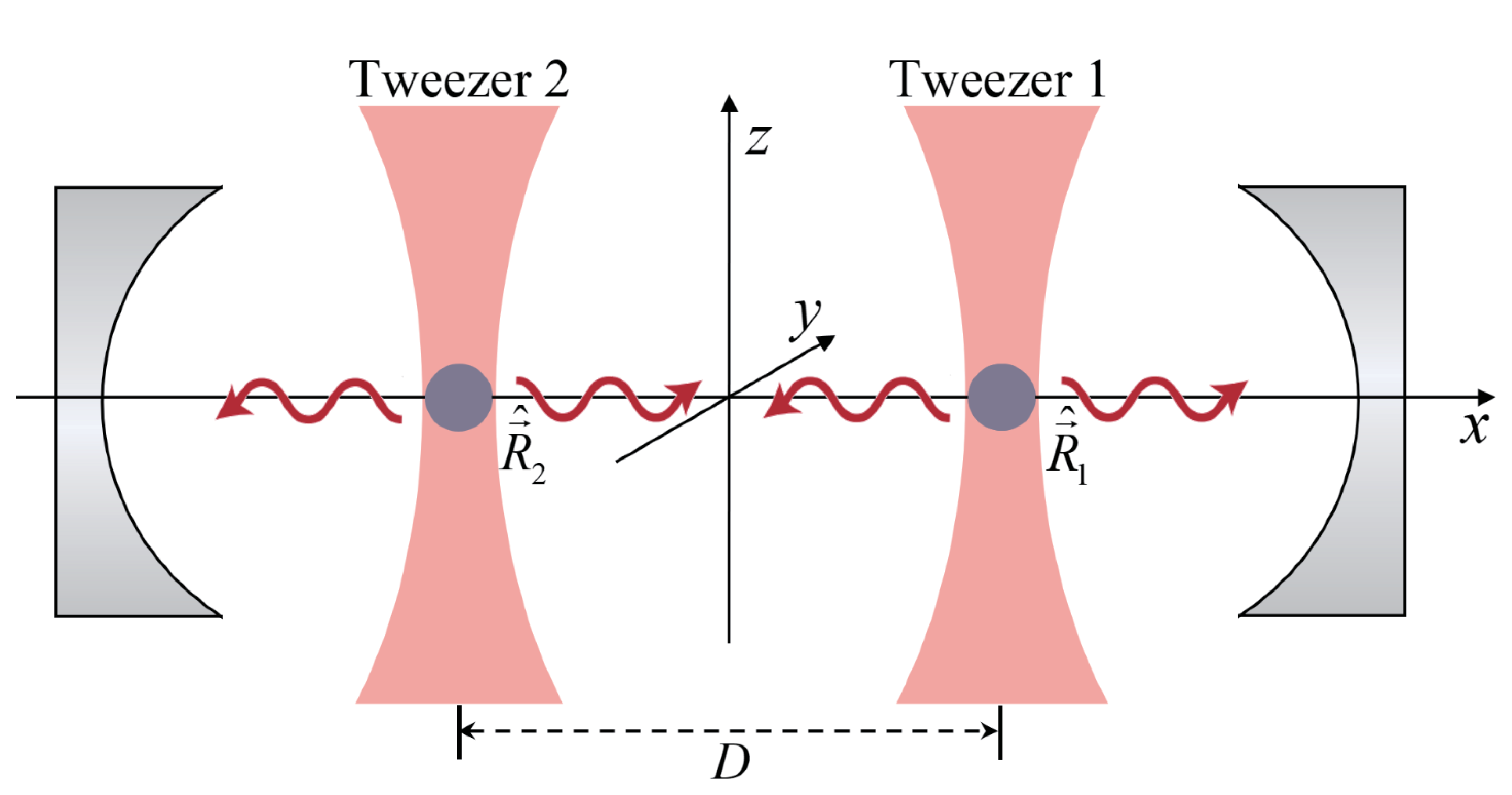}
\caption{{Schematic of the physical setup. Two dielectric nanoparticles are
trapped at positions $\boldsymbol{\hat{R}}_{1}$
and $\boldsymbol{\hat{R}}_{2}$ by two optical tweezers, where the cavity axis is along the $x$ direction, and the two optical tweezers propagate along the $z$ axis with polarization along the $y$ direction. The distance between the foci of the two tweezers is $D$.}}
\label{modelv1}
\end{figure}

We consider a coupled cavity-levitated-nanoparticle system, in which two dielectric nanoparticles trapped by two optical tweezers are coupled to the field modes in a Fabry-P\'{e}rot cavity, as shown in Fig.~\ref{modelv1}. The Fabry-P\'{e}rot cavity, with the cavity axis aligning with the $x$ direction, contains two nanoparticles. The two nanoparticles, placed at $\boldsymbol{\hat{R}}_{1}=(\hat{X}_{1},\hat{Y}_{1},\hat{Z}_{1})$
and $\boldsymbol{\hat{R}}_{2}=(\hat{X}_{2},\hat{Y}_{2},\hat{Z}_{2})$, have the radius $a_{0}=90$ nm, density $\rho\approx2200 $ kg/$m^{3}$,
and dielectric constant $\epsilon _{r}=2.07$. We assume that the two
optical tweezers have electric fields propagating along the $z$ axis, with the corresponding polarizations $\boldsymbol{e}_{%
\text{tw}}^{(1)}$ and $\boldsymbol{e}_{\text{tw}}^{(2)}$ along the $y$ direction. The foci of the two optical tweezers
are located at the positions $\left( x_{10},0,0\right) $ and $%
\left( x_{20},0,0\right) $, separated by a distance $D$. The frequencies of the
two optical tweezers are $\omega _{\text{tw}}=2\pi c/\lambda _{\text{tw}}$,
where $c$ is the speed of light in a vacuum and $\lambda _{\text{tw}}$ is the
wavelength of the tweezers.

%%%%%%%%%%%%%%%%%%%%%%%%%%%%%
%%%%%%%%%%%%%%%%%%%%%%%%%%%%%
The total Hamiltonian of the system can be written as%
\begin{equation}
\hat{H}_{\text{tot}}=\hat{H}_{\text{np}}+\hat{H}_{\text{cav}}+\hat{H}_{\text{int}}.  \label{1}
\end{equation}%
Here, the Hamiltonian $\hat{H}_{\text{np}}$ describes the kinetic energy of the center-of-mass
motion for the two nanoparticles, and it takes the form
\begin{equation}
\hat{H}_{\text{np}}=\sum_{j=1,2}\frac{\hat{\mathbf{P}}_{j}^{2}}{2m},
\end{equation}%
where $\hat{\mathbf{P}}_{j}=( \hat{P}_{jx},\hat{P}_{jy},\hat{P}_{jz}) $ is the three-dimensional momentum
operator for the $j$th $\left( j=1,2\right) $ nanoparticle with mass $m$. The second term on the right-hand side of Eq.~(\ref{1}) reads
\begin{eqnarray}
\hat{H}_{\text{cav}}&=&\frac{1}{2}\int [ \varepsilon _{0}\boldsymbol{E}_{%
\text{cav}}^{2}\left( \boldsymbol{r}\right) +\boldsymbol{B}%
_{\text{cav}}^{2}\left( \boldsymbol{r}\right)/\mu _{0} ] d\boldsymbol{r} \notag \\
&=&\sum_{j}\hbar \omega _{j}(\hat{a}_{j}^{\dagger }\hat{a}_{j}+1/2)\text{,}
\end{eqnarray}
where $\boldsymbol{E}_{\text{cav}}$ and $\boldsymbol{B}_{\text{cav}}$ are, respectively, the electric and magnetical fields in the cavity, and $\varepsilon _{0}$ $(\mu _{0})$ is the free space premittivity (permeability). In addition, $\omega _{j}$ is the resonance frequency of the $j$th cavity mode (described by the creation and annihilation operators $\hat{a}_{j}^{\dagger}$ and $\hat{a}_{j}$) in the optical cavity.
Since the frequency of the center-of-mass motion for the nanoparticles
is much smaller than the free spectrum range of the cavity, we could consider that the two nanoparticles
are coupled to a single cavity field mode. Then, the Hamiltonian of the optical cavity can be approximately denoted as $\hat{H}_{\text{cav}}\approx\hbar\omega_{\text{cav}}\hat{a}^{\dagger}\hat{a}$, where $\omega_{\text{cav}}$ is
the resonance frequency of the cavity mode under consideration with the wave number $k$, described by the creation and annihilation operators $ \hat{a}^{\dagger}$ and $\hat{a}$. Note that the zero-point fluctuation term $\hbar\omega_{\text{cav}}/2$ has been omitted in the Hamiltonian.

The last term $\hat{H}_{\text{int}}$ in Eq.~(\ref{1}) describes the
interactions between the nanoparticles and the electromagnetic fields. In
the Rayleigh regime, the radius of the nanoparticle is much smaller than the
optical wavelength ($a_{0}\ll $ $\lambda $), and the interaction Hamiltonian between the nanoparticles and the electric fields can be
written as~\cite{UPRL2019,DPRL2019,CPRA2019}%
\begin{equation}
\hat{H}_{\text{int}}\approx -%
\frac{1}{2}\sum_{j=1,2}\alpha\boldsymbol{E}^{2}( \boldsymbol{\hat{R}}_{j}),  \label{3}
\end{equation}%
where $\alpha=\varepsilon _{0}\varepsilon _{c}V$ is the particle polarizability with $\varepsilon _{c}=3(\epsilon _{r}-1)/(\epsilon _{r}+2)$ and $V$ being the volume of the nanoparticle. %
In Eq.~(\ref{3}), $\boldsymbol{E}(\boldsymbol{\hat{R}}_{j})$ represents the electric field at the position of the $j$th particle,
where $\boldsymbol{\hat{R}}_{j}=\boldsymbol{r}_{j0}+%
\boldsymbol{\hat{r}}_{j}$ denotes the center-of-mass position operator of
the $j$th particle, with $\boldsymbol{r}_{j0}=( x_{j0},0,0) $
being the focus of the $j$th optical tweezer along the cavity axis
and $\boldsymbol{\hat{r}}_{j}=(\hat{x}_{j},\hat{y}_{j},\hat{z}_{j})$ the
position operator of the $j$th particle.
\subsection{The initial electric field}
In general, the total electric field at the position $\boldsymbol{r}=\left(
x,y,z\right) $ can be approximately written as a sum of the
cavity field $\boldsymbol{\hat{E}}_{\text{cav}}$ and the
fields $\bm{\mathcal{E}}_{\text{tw}}^{(j)}$ for the two optical tweezers,%
\begin{equation}
\boldsymbol{\hat{E}}_{I}\left( \boldsymbol{r}\right) =\boldsymbol{\hat{E}}_{\text{cav}%
}\left( \boldsymbol{r}\right) +\sum_{j=1,2}\bm{\mathcal{E}}_{\text{tw}%
}^{(j)}\left( \boldsymbol{r}\right) \text{.} \label{20}
\end{equation}%
The first term on the right-hand side of Eq.~(\ref{20}) describes the single-mode electric field of the cavity, which is given by%
\begin{equation}
\boldsymbol{\hat{E}}_{\text{cav}}\left( \boldsymbol{r}\right) =\epsilon _{\text{cav%
}}\cos \left( kx-\phi \right) ( \hat{a}^{\dagger }+\hat{a}) \boldsymbol{e}_{%
\text{cav}},  \label{5}
\end{equation}%
where $\epsilon _{\text{cav}}=\sqrt{\hbar \omega _{\text{cav}}/({%
2\varepsilon _{0}V_{\text{cav}}})}$ is the amplitude at the center of the cavity with $V_{\text{cav}}$ being the cavity volume. For simplicity, we consider the case $\phi =0$ and the $y$-axis polarized cavity mode in this paper.

We assume that the two tweezers are sufficiently spaced apart such that the influence of the electric
field $\bm{\mathcal{E}}_{\text{tw1}}$ of the tweezer 1 on the distant nanoparticle 2 can be neglected,
and vice versa. Then the total electric fields at the positions of the two nanoparticles 1 and 2 can be approximated as%
\begin{subequations}
\begin{align}
\boldsymbol{\hat{E}}_{I}^{(1)}( \boldsymbol{\hat{R}}_{1})  &=%
\bm{\mathcal{E}}_{\text{tw$1$}}( \boldsymbol{\hat{R}}_{1}) +\boldsymbol{\hat{E}}%
_{\text{cav}}( \boldsymbol{\hat{R}}_{1}),  \\
\boldsymbol{\hat{E}}_{I}^{(2)}( \boldsymbol{\hat{R}}_{2})  &=%
\bm{\mathcal{E}}_{\text{tw$2$}}( \boldsymbol{\hat{R}}_{2}) +\boldsymbol{\hat{E}}%
_{\text{cav}}( \boldsymbol{\hat{R}}_{2}).
\end{align}
\end{subequations}%
Typically, the fields of the optical tweezers are considered in
 coherent states and thus they can be well described by classical fields. Then the electric field of the $j$%
th optical tweezer can be expressed by $\bm{\mathcal{E}}_{\text{tw}j}\left(
\boldsymbol{r},t\right)=\text{Re}[\boldsymbol{E}_{\text{tw}j}\left(
\boldsymbol{r},t\right)]$, where
\begin{equation}
\boldsymbol{E}_{\text{tw}j}\left(
\boldsymbol{r},t\right) =E_{j0}\left(
\boldsymbol{r}\right) e^{-i\left[ k_{\text{tw}}z+\phi _{\text{t}}\left(
\boldsymbol{r}\right) \right] }e^{-i\omega _{\text{tw}}t}\boldsymbol{e}_{\text{tw}}^{(j)},  \label{4}
\end{equation}%
with the laser frequency $\omega _{\text{tw}}=ck_{\text{tw}}$ and
wave number $k_{\text{tw}}=2\pi /\lambda _{\text{tw}}$ of the tweezer.
We assume that the propagating directions of the two beams are parallel, and that
the polarizations $\boldsymbol{e}_{\text{tw}}^{(1)}$ and $\boldsymbol{e}_{\text{tw}}^{(2)}$ of the electric fields are along the $y$ direction. Then the real amplitude $E_{j0}\left( \boldsymbol{r}\right) $ in Eq.~(\ref{4}) can be written as
\begin{equation}
E_{j0}\left( \boldsymbol{r}\right) =\epsilon _{\text{tw}}^{(j)}\frac{1}{\sqrt{%
1+\left( z/z_{R}\right) ^{2}}}\exp\left(-\frac{( x-x_{j0}) ^{2}+y^{2}}{%
W^{2}\left( z\right) }\right)\text{,} \label{Ej0}
\end{equation}%
where $\epsilon _{\text{tw}}^{(j)}=\sqrt{4P_{\text{tw}}^{(j)}/({\pi
\varepsilon _{0}cW_{t}^{2}})}$ is the amplitude of the electric field, with $P_{\text{tw}}^{(j)}$ being the power of the $j$th laser and $W_{t}$ the tweezer waist at the focus. In addition, $%
z_{R}=\pi W_{t}^{2}/\lambda$ is the Rayleigh range and
$W\left( z\right) =W_{t}\sqrt{1+\left( z/{z_{R}}\right)
^{2}}$. Note that the phase factor $\phi _{\text{t}}\left( \boldsymbol{r}%
\right) \approx \arctan \left( z/z_{R}\right) -k_{\text{tw}}z%
[( x-x_{j0})^{2} +y^{2}]/(2z^{2}+2z_{R}^{2})$ in Eq.~(\ref{4}) can be neglected,
since the Rayleigh range $z_{R}$ is typically
several orders of magnitude larger than other length scales. %
It should be pointed out that, for realistic cases, the phase difference between the two tweezers is a crucial manipulation means, which can be used to control the couplings between the two particles. In this work, we consider a simple case by choosing a zero phase difference, which has been implemented in experiments~\cite{JS2022}. For a non-zero phase difference, a direct coupling between the $x$-direction and $z$-direction motional modes will be induced, and hence it will make the coupling configuration complicated.

\subsection{The radiation fields}
In the Rayleigh regime, the nanoparticle embedded in the electric fields will
possess an electric dipole moment, which will create electromagnetic radiation
by charge oscillation. Physically, the frequency of the radiation field is equal to that of the
incident field. Below, we derive the relation between the radiation field and the dipole. For simplicity in the derivation of the general relation, we use $\boldsymbol{r}_{1}$ and $\boldsymbol{r}_{2}$ $($different from $\boldsymbol{\hat{r}}_{j}$ introduced before$)$ to denote the positions of the considered point and the dipole, respectively. The electric field at position $\boldsymbol{r}_{1}$ generated by the oscillating dipole at the position $\boldsymbol{r}_{2}$ is given by
\begin{equation}
\boldsymbol{E}%
_{\text{rad}}\left( \boldsymbol{r}_{1}\right) =\overleftrightarrow{\mathbf{G}}\left( \boldsymbol{r}_{1}-%
\boldsymbol{r}_{2}\right)\cdot \boldsymbol{P}\left( \boldsymbol{r}_{2}\right) ,
\end{equation}%
where the field propagator (also know as the dyadic Green's function)
between the two dipoles is given by~\cite{KRMP2010,Lbook2012,MPRA2018}
\begin{eqnarray}
\overleftrightarrow{\mathbf{G}}\left( \boldsymbol{r}_{1}-\boldsymbol{r}_{2}\right)  &=&\frac{%
e^{ik_{0}r_{0}}}{4\pi \varepsilon _{0}r_{0}}\left[ \left( \frac{1-ik_{0}r_{0}}{%
r_{0}^{2}}\right) \frac{3\boldsymbol{r}_{0}\boldsymbol{r}_{0}
-r_{0}^{2}}{r_{0}^{2}}\right.   \notag \\
&&\left. +k_{0}^{2}\frac{r_{0}^{2}-\boldsymbol{r}_{0}\boldsymbol{r}%
_{0}}{r_{0}^{2}}\right].   \label{15}
\end{eqnarray}%
In Eq.~(\ref{15}), $k_{0}$ is the wave number of the incident field, $r_{0}=\left\vert \boldsymbol{r}_{0}\right\vert =\left\vert \boldsymbol{%
r}_{1}-\boldsymbol{r}_{2}\right\vert $ is the distance between the two
dipoles. Based on Eq.~(\ref{15}), we have the relation $\overleftrightarrow{\mathbf{G}}\left( \boldsymbol{r}_{1}-\boldsymbol{r}%
_{2}\right) =\overleftrightarrow{\mathbf{G}}\left( \boldsymbol{r}_{2}-\boldsymbol{r}_{1}\right) =%
\overleftrightarrow{\mathbf{G}}\left( \boldsymbol{r}_{0}\right) $, which can be used to describe the fields realted to the two particles. In addition, $\boldsymbol{r}_{0}\boldsymbol{r}_{0}=\sum_{QQ^{\prime }}Q_{0}Q_{0}^{\prime }\boldsymbol{e}_{Q}\boldsymbol{e}_{Q^{\prime }}$ with $\boldsymbol{e}_{Q}$ being the unit vector in the $Q$ direction and $Q,Q^{\prime }=x,y,z$. In the following analyses,
the Green function can be divided into two parts: $\alpha\overleftrightarrow{\mathbf{G}}\left( \boldsymbol{r}_{0}\right) =e^{ik_{0}r_{0}}%
[ \eta _{n}\left(D/r_{0}\right)^{3}\left( 1-ik_{0}r_{0}\right) \overleftrightarrow{\mathbf{M}}_{n}\left( \boldsymbol{r}%
_{0}\right) +\eta _{f}\left(D/r_{0}\right)\overleftrightarrow{\mathbf{M}}_{f}\left( \boldsymbol{r}_{0}\right) ]
$~\cite{Daxv2203}, where $\eta _{n}=1 /4\pi \varepsilon_{0}D^{3}$ is the near-field constant and $\eta _{f}= k_{0}^{2}/4\pi \varepsilon _{0}D$
is the far-field constant. The near-field constant $\eta _{n}$ is much smaller than the far-field constant $%
\eta _{f}$ in the far-field regime $k_{0}r_{0}\gg 1$. In addition, we introduce the near-field tensor%
\begin{eqnarray}
\overleftrightarrow{\mathbf{M}}_{n}\left( \boldsymbol{r}\right)&=&\frac{1}{r^{2}}[( 3x^{2}-r^{2}) \boldsymbol{e}_{x}%
\boldsymbol{e}_{x}+3xy\boldsymbol{e}_{x}\boldsymbol{e}_{y}+3xz\boldsymbol{e}%
_{x}\boldsymbol{e}_{z} \notag\\
&&+3xy\boldsymbol{e}_{y}\boldsymbol{e}_{x}+( 3y^{2}-r^{2})
\boldsymbol{e}_{y}\boldsymbol{e}_{y}+3yz\boldsymbol{e}_{y}\boldsymbol{e}_{z}
\notag\\
&&+3xz\boldsymbol{e}_{z}\boldsymbol{e}_{x}+3yz\boldsymbol{e}_{z}\boldsymbol{e%
}_{y}+( 3z^{2}-r^{2}) \boldsymbol{e}_{z}\boldsymbol{e}_{z}],
\end{eqnarray}%
and the far-field tensor
\begin{eqnarray}
\overleftrightarrow{\mathbf{M}}_{f}\left( \boldsymbol{r}\right)&=&\frac{1}{r^{2}}[( r^{2}-x^{2}) \boldsymbol{e}_{x}\boldsymbol{%
e}_{x}-xy\boldsymbol{e}_{x}\boldsymbol{e}_{y}-xz\boldsymbol{e}_{x}%
\boldsymbol{e}_{z} \notag\\
&&-xy\boldsymbol{e}_{y}\boldsymbol{e}_{x}+( r^{2}-y^{2})
\boldsymbol{e}_{y}\boldsymbol{e}_{y}-yz\boldsymbol{e}_{y}\boldsymbol{e}_{z}
\notag\\
&&-xz\boldsymbol{e}_{z}\boldsymbol{e}_{x}-yz\boldsymbol{e}_{z}\boldsymbol{e}%
_{y}+( r^{2}-z^{2}) \boldsymbol{e}_{z}\boldsymbol{e}_{z}].
\end{eqnarray}%

For our considered coupled cavity-levitated-particle system, the total electric field for the $j$th ($j=1,2$) particle is given by the sum of the
incident field $\boldsymbol{E}_{I}^{(j)}$ and the field emitted by the other
dipole,%
\begin{eqnarray}
\boldsymbol{E}_{\text{tot}}^{(j)}( \boldsymbol{\hat{R}}_{j})  &=&%
\boldsymbol{E}_{I}^{(j)}( \boldsymbol{\hat{R}}_{j}) +\overleftrightarrow{\mathbf{G}}(
\boldsymbol{\hat{R}}_{0})\cdot \boldsymbol{P}^{(\bar{j})}( \boldsymbol{\hat{R}}_{\bar{j}}) \text{.}
\end{eqnarray}%
Here, $\boldsymbol{\hat{R}}_{0}=\boldsymbol{\hat{R}}_{1}-\boldsymbol{\hat{R}}_{2}=(\hat{X}_{0},\hat{Y}_{0},\hat{Z}_{0})$ and $\boldsymbol{P}^{(\bar{j})}( \boldsymbol{\hat{R}}_{\bar{j}}) =\alpha%
\boldsymbol{E}_{\text{tot} }^{(\bar{j})}( \boldsymbol{\hat{R}}%
_{\bar{j}}) $ is the dipole moment generated by the $\bar{j}$th dielectric nanoparticle, where the index $\bar{j}$ denotes the other particle with respect to the $j$th particle (namely $\bar{1}=2$ and $\bar{2}=1$).
Since the cavity field and the tweezer field have
different wave numbers, the Green function will take two distinct forms,
$\overleftrightarrow{\mathbf{G}}_{\text{cav}}$ and $\overleftrightarrow{\mathbf{G}}_{\text{tw}}$, corresponding to the wave numbers of the cavity field and the tweezer field, respectively. Consequently, the electric field $\boldsymbol{E}_{%
\text{tot}}^{(j)}(\boldsymbol{\hat{R}}_{j})$ can be divided into two parts, $\boldsymbol{\hat{E}}_{
\text{tot}}^{(j)}(\boldsymbol{\hat{R}}_{j})=\boldsymbol{E}_{\text{tot,tw}}^{(j)}(\boldsymbol{\hat{R}}_{j})+\boldsymbol{E}_{\text{tot,cav}}^{(j)}(\boldsymbol{\hat{R}}_{j})$, with
\begin{subequations}
\label{26}
\begin{align}
\boldsymbol{E}_{\text{tot,tw}}^{(j)}( \boldsymbol{\hat{R}}_{j})  &\approx%
\boldsymbol{E}_{\text{tw}j}( \boldsymbol{\hat{R}}_{j},t) +\overleftrightarrow{\mathbf{G}}_{\text{tw}}(
\boldsymbol{\hat{R}}_{0})\cdot \alpha\boldsymbol{E}_{\text{tot,tw}}^{(\bar{j})}( \boldsymbol{\hat{R}}_{\bar{j}}) \text{,}
\\
\boldsymbol{\hat{E}}_{\text{tot,cav}}^{(j)}( \boldsymbol{\hat{R}}_{j}) &\approx\boldsymbol{\hat{E}}_{\text{cav}}( \boldsymbol{\hat{R}}_{j}) +\overleftrightarrow{\mathbf{G}}_{\text{cav}}(\boldsymbol{\hat{R}}_{0})\cdot \alpha\boldsymbol{\hat{E}}_{\text{tot,cav}}^{(\bar{j})}( \boldsymbol{\hat{R}}_{\bar{j}})   \text{.}
\end{align}
\end{subequations}%
Since the magnitude of $\alpha\overleftrightarrow{\mathbf{G}}$ is considerably small
compared to the trapping fields, we can neglect the second-order terms in Eqs.~(\ref{26}). Then we
obtain the total electric field consisting of the incident field [the tweezer field $%
\bm{\mathcal{E}}_{\text{tw}}( \boldsymbol{\hat{R}},t) $ and the cavity field $%
\boldsymbol{\hat{E}}_{\text{cav}}( \boldsymbol{\hat{R}}) $] and the emitted
field [$\bm{\mathcal{E}}_{\text{Gtw}}( \boldsymbol{\hat{R}}) $ and $%
\bm{\mathcal{\hat{E}}}_{\text{Gcav}}( \boldsymbol{\hat{R}}) $] by the dipole,
\begin{eqnarray}
\bm{\mathcal{\hat{E}}}_{\text{tot}}^{(j)}( \boldsymbol{\hat{R}}_{j})  &=&%
\bm{\mathcal{E}}_{\text{tw}j}( \boldsymbol{\hat{R}}_{j},t) +\boldsymbol{\hat{E}}%
_{\text{cav}}( \boldsymbol{\hat{R}}_{j})   \notag \\
&&+\bm{\mathcal{E}}_{\text{Gtw}\bar{j}}( \boldsymbol{\hat{R}}_{j}) +%
\bm{\mathcal{\hat{E}}}_{\text{Gcav}}( \boldsymbol{\hat{R}}_{j}),   \label{2}
\end{eqnarray}%
where $\bm{\mathcal{E}}_{\text{Gtw}\bar{j}}( \boldsymbol{\hat{R}}_{j}) =%
\text{Re}[\overleftrightarrow{\mathbf{G}}_{\text{tw}}( \boldsymbol{\hat{R}}_{0})\cdot \alpha%
\boldsymbol{E}_{\text{tw}\bar{j}}( \boldsymbol{\hat{R}}_{\bar{j}},t)] $ describes the
radiation field generated by the oscillating dipoles at $\boldsymbol{\hat{R}}%
_{\bar{j}}$, which is induced by the $ \bar{j} $th tweezer field $\boldsymbol{E}%
_{\text{tw}\bar{j}}( \boldsymbol{\hat{R}}_{\bar{j}}) $. In addition, $\bm{\mathcal{\hat{E}}}_{%
\text{Gcav}}( \boldsymbol{\hat{R}}_{j}) =\text{Re}[\overleftrightarrow{\mathbf{G}}_{\text{cav}}( \boldsymbol{\hat{R}%
}_{0})\cdot\alpha\boldsymbol{\hat{E}}_{\text{cav}}( \boldsymbol{\hat{R}}%
_{\bar{j}})] $ represents the radiation field produced by the dipole moment, which is
induced by the cavity field at the position $\boldsymbol{\hat{R}}_{\bar{j}}$.

\subsection{The interaction Hamiltonians}
In this section, we present the detailed expressions of the interaction Hamiltonians by putting the electric
field operator $\bm{\mathcal{E}}_{\text{tot}}^{(j)}( \boldsymbol{\hat{R}}_{j},t)$ given by Eq.~(\ref{2})
into Hamiltonian~(\ref{3}). The interaction Hamiltonian can be divided
into two parts $\hat{H}_{\text{int}}=\sum_{j=1,2}\hat{H}_{\text{int}%
}^{(j)}$, where the forms of the two parts are similar.
Below, we take the $j$th (for $j$=1,2) particle as an example. The Hamiltonian $\hat{H}_{\text{int}}^{(j)}$ can be written as
\begin{eqnarray}
\hat{H}_{\text{int}}^{(j)} &=&-\frac{1}{2}\alpha [ \mathbf{\bm{\mathcal{E}}}_{\text{tw}%
j}( \boldsymbol{\hat{R}}_{j},t) +\boldsymbol{\hat{E}}_{\text{cav}}(
\boldsymbol{\hat{R}}_{j})   \notag \\
&& +\bm{\mathcal{E}}_{\text{Gtw}\bar{j}}( \boldsymbol{\hat{R}}_{j}) +%
\bm{\mathcal{\hat{E}}}_{\text{Gcav}}( \boldsymbol{\hat{R}}_{j}) ] ^{2}\text{,}
\end{eqnarray}%
which can be further divided into six terms%
\begin{eqnarray}
\hat{H}_{\text{int}}^{(j)} &=&\hat{H}_{\text{cs}}^{(j)}+\hat{H}_{\text{rad-rad}}^{(j)}+\hat{H}_{\text{%
tw-Gtw}}^{(j)} \notag\\
&&+\hat{H}_{\text{cav-Gcav}}^{(j)}+\hat{H}_{\text{tw-Gcav}}^{(j)}+\hat{H}_{\text{cav-Gtw}}^{(j)}, \label{25}
\end{eqnarray}
each of which represents a special physical interaction.

The first term $\hat{H}_{\text{cs}}^{(j)}$ in Eq.~(\ref{25}) is the standard interaction Hamiltonian of the cavity-field with the $j$th
levitated particle
\begin{equation}
\hat{H}_{\text{cs}}^{(j)}=-\frac{1}{2}\alpha [ \bm{\mathcal{E}}_{\text{tw}j}(
\boldsymbol{\hat{R}}_{j},t) +\boldsymbol{\hat{E}}_{\text{cav}}( \boldsymbol{\hat{R}}%
_{j}) ] ^{2}, \label{Hcs}
\end{equation}%
which consists of three parts $\hat{H}_{\text{cs}}^{(j)}=\hat{H}_{\text{tw-tw}}^{(j)}+\hat{H}_{\text{cav-cav}}^{(j)}+\hat{H}_{\text{cav-tw}}^{(j)}$.
Since the $j$th particle is trapped near the focus of the $j$th tweezer, we can
approximate the electric field of the tweezer by its expansion near $%
\boldsymbol{r}_{j0}=( x_{j0},0,0) $. Then, we can obtain the harmonic potential energy of the tweezer as
$\hat{H}_{\text{tw-tw}}^{(j)}=-\alpha \bm{\mathcal{E}}_{\text{tw}j}^{2}(
\boldsymbol{\hat{R}}_{j})/2\approx \sum_{Q} m\omega_{jQ}^{2}\hat{Q}_{j}^{2}/2$
with $ Q=x,y,z$, where we employ the rotating-wave approximation
and neglect both the $\exp \left( \pm 2i\omega _{\text{tw}}t\right) $ terms and the
constant terms. This means that the $j$th nanoparticle is trapped by the tweezer with trapping frequencies $[
\omega _{jx},\omega _{jy},\omega _{jz}] =\sqrt{\alpha /2m}\epsilon _{%
\text{tw}}^{(j)}[ \sqrt{2}W_{t}^{-1},\sqrt{2}W_{t}^{-1},z_{R}^{-1}] $.
The square term of the cavity field contains both the cavity frequecy shift and the radiation
pressure effect, $\hat{H}_{\text{cav-cav}}^{(j)}=-\alpha \boldsymbol{\hat{E}}_{\text{cav}}^{2}(\boldsymbol{\hat{R}}_{j})/2\approx\hbar
\omega_{\text{sh}}\hat{a}^{\dagger }\hat{a}+\hbar g_{ax_{j}}\hat{a}^{\dagger }\hat{a}\hat{x}_{j}$, where $\omega_{\text{sh}}=-\alpha \epsilon_{\text{cav}}^{2}\cos ^{2}
( kx_{j0})/\hbar$ and $g_{ax_{j}}=\alpha \epsilon _{\text{cav}}^{2}2\cos ( kx_{j0}) \sin (kx_{j0}) k/\hbar$. In addition, the interaction term between the tweezer and cavity fields is given by $\hat{H}_{\text{tw-cav}}^{(j)}=-\alpha[
\bm{\mathcal{E}}_{\text{tw}j}(\boldsymbol{\hat{R}}_{j})\cdot\boldsymbol {\hat{E}}_{\text{cav}}(\boldsymbol{\hat{R}}_{j})]
\approx \hbar \Omega ( \hat{a}^{\dagger }+\hat{a})+\hbar g_{x_{j}} ( \hat{a}^{\dagger }+\hat{a}) \hat{x}_{j}+i\hbar g_{z_{j}}(\hat{a}-\hat{a}^{\dagger }) \hat{z}_{j}$, which describes
the displacement of the cavity mode and the coupling mediated by coherent scattering, where $\Omega=-\alpha \epsilon _{\text{cav}}\epsilon _{\text{tw}}^{(j)}\cos
( kx_{j0})/(2\hbar) $, $g_{x_{j}}=\alpha \epsilon _{\text{cav}}\epsilon _{\text{tw}}^{(j)}\sin
( kx_{j0})k/(2\hbar)$, and $g_{z_{j}}=-\alpha \epsilon _{\text{cav}}\epsilon _{\text{tw}}^{(j)}\cos
( kx_{j0}) k_{\text{tw}}/(2\hbar)$~\cite{UPRL2019,DPRL2019,CPRA2019}.

The second term $\hat{H}_{\text{rad-rad}}^{(j)}$  in Eq.~(\ref{25}) describes the
interaction between these radiation fields at position $%
\boldsymbol{\hat{R}}_{j}$ generated by the $\bar{j}$th oscillating dipole, and it is written as
\begin{equation}
\hat{H}_{\text{rad-rad}}^{(j)}=-\frac{1}{2}\alpha [ \bm{\mathcal{E}}_{\text{Gtw$\bar{j}$}%
}( \boldsymbol{\hat{R}}_{j}) +\bm{\mathcal{\hat{E}}}_{\text{Gcav}}(
\boldsymbol{\hat{R}}_{j}) ] ^{2}. \label{Hrr}
\end{equation}%
We point out that the terms $\hat{H}_{\text{rad-rad}}^{(j)}$ for $j=1,2$ are higher-order term of $\alpha$ compared to the other terms, so the terms are usually small enough to be ignored.

The remaining terms in Eq.~(\ref{25}) describe the interactions between the
incident field at the position of $j$th particle and the field at
position $\boldsymbol{\hat{R}}_{j}$ generated by the $\bar{j}$th dipole. Concretely, these interaction Hamiltonians are given by
\begin{subequations}
\label{crossterm}
\begin{align}
\hat{H}_{\text{tw-Gtw}}^{(j)} &=-\alpha \bm{\mathcal{E}}_{\text{tw}j}(
\boldsymbol{\hat{R}}_{j},t) \cdot \bm{\mathcal{E}}_{\text{Gtw}\bar{j}}(
\boldsymbol{\hat{R}}_{j}) ,  \label{12} \\
\hat{H}_{\text{cav-Gcav}}^{(j)} &=-\alpha \boldsymbol{\hat{E}}_{\text{cav}}(
\boldsymbol{\hat{R}}_{j}) \cdot \bm{\mathcal{\hat{E}}}_{\text{Gcav}}( \boldsymbol{%
\hat{R}}_{j}) ,  \label{13} \\
\hat{H}_{\text{tw-Gcav}}^{(j)} &=-\alpha \bm{\mathcal{E}}_{\text{tw}j}(
\boldsymbol{\hat{R}}_{j},t) \cdot \bm{\mathcal{\hat{E}}}_{\text{Gcav}}(
\boldsymbol{\hat{R}}_{j}) , \label{Htgc} \\
\hat{H}_{\text{cav-Gtw}}^{(j)} &=-\alpha \boldsymbol{\hat{E}}_{\text{cav}}(
\boldsymbol{\hat{R}}_{j}) \cdot \bm{\mathcal{E}}_{\text{Gtw}\bar{j}}(
\boldsymbol{\hat{R}}_{j}) . \label{Hcgt}
\end{align}
\end{subequations}%
The four cross terms describe
the interactions between the incident field and the radiation field, and they
are generated by the mechanical effect of scattered light via the optical binding force. Equations~(\ref{12}) and~(\ref{13}) describe the
lateral binding and longitudinal binding, respectively. The optical binding force can be
calculated as~\cite{KRMP2010,JS2022}
\begin{equation}
\boldsymbol{F}^{\text{bind}}=\frac{1}{2}\nabla \mathbf{\text{Re}}\left[ \boldsymbol{P}%
^{\ast }( \boldsymbol{R}_{j})\cdot \overleftrightarrow{\mathbf{G}}( \boldsymbol{R}_{j}-%
\boldsymbol{R}_{i})\cdot \alpha \boldsymbol{E}\left( \boldsymbol{R%
}_{i}\right) \right],   \label{14}
\end{equation}%
where the force term describes the interaction between the emitted field
and the dipole at $\boldsymbol{R}_{j}$. Equation~(\ref{12}) describes the interaction corresponding to the case where the two
particles are placed on the $x$-axis, and they are trapped by the two
optical tweezers with the same frequency, respectively. Here, the two tweezers polarize along the $y$-axis. In this
case, the binding force acting on particle 1 has the following components:
\begin{subequations}
\label{31}
\begin{align}
F_{x}^{(1)} =&\frac{\alpha ^{2}E_{10}E_{20}}{8\pi \epsilon _{0}R_{0}^{4}}[3\cos
\left( k_{\text{tw}}R_{0}\right) +3k_{\text{tw}}R_{0}\sin \left( k_{\text{tw}%
}R_{0}\right)  \notag \\
&-2\left( k_{\text{tw}}R_{0}\right) ^{2}\cos \left( k_{\text{tw}%
}R_{0}\right) -\left( k_{\text{tw}}R_{0}\right) ^{3}\sin \left( k_{\text{tw}%
}R_{0}\right) ], \\
F_{y}^{(1)} =&0, \\
F_{z}^{(1)} =&\frac{\alpha ^{2}E_{10}E_{20}}{8\pi \epsilon _{0}R_{0}^{4}}[-k_{%
\text{tw}}R_{0}\sin \left( k_{\text{tw}}R_{0}\right)   \notag \\
&+\left( k_{\text{tw}}R_{0}\right) ^{2}\cos \left( k_{\text{tw}%
}R_{0}\right) +\left( k_{\text{tw}}R_{0}\right) ^{3}\sin \left( k_{\text{tw}%
}R_{0}\right) ],
\end{align}
\end{subequations}%
along the $x$, $y$, and $z$ axes, respectively, with the distance $R_{0}$ between the two particles.

\begin{figure}[t]
\center\includegraphics[width=0.48\textwidth]{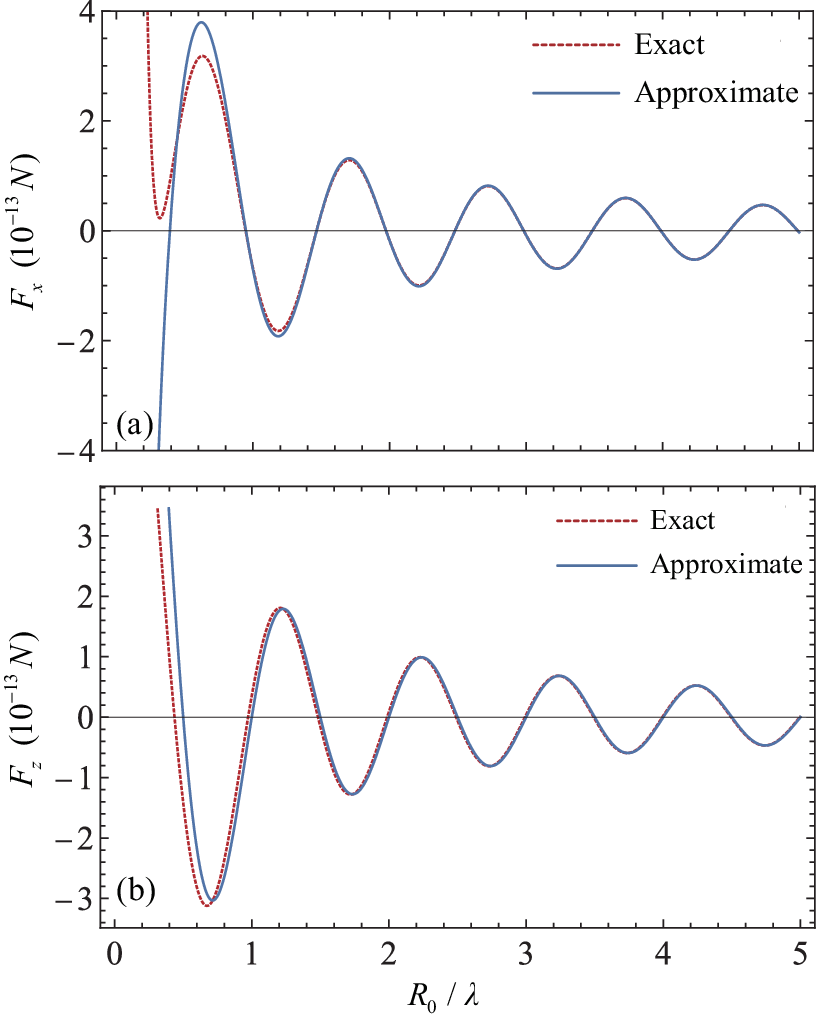}
\caption{{Comparison of the exact and approximate results concerning the optical binding forces (a) $F_{x}$ and (b) $F_{z}$ (along the $x$ axis and $z$ axis) between the two nanoparticles as
functions of the scaled distance $R_{0}/\lambda$. The radius of the two nanoparticles is $a_{0}=90$ nm, the numerical apertures of the two tweezers are  $\text{NA}\approx0.8$, and the powers of the two tweezers are $P_{\text{tw}}^{(1)}=0.8$ W and $P_{\text{tw}}^{(2)}=0.45$ W.}}
\label{modelv2}
\end{figure}

The Green function $\overleftrightarrow{\mathbf{G}}(r_{0})$ [Eq.~(\ref{15})] contains these terms proportional to $r_{0}^{-1}$, $r_{0}^{-2}$, and $r_{0}^{-3}$. In the far-field region $%
k_{0}r_{0}\gg 1$, the terms with $r_{0}^{-1}$ dominate in the case, and then the Green function retains only the last term, i.e., $\alpha\overleftrightarrow{\mathbf{G}}\left( \boldsymbol{r}_{0}\right)\approx e^{ik_{0}r_{0}}\eta _{f}\left(D/r_{0}\right)\overleftrightarrow{\mathbf{M}}_{f}\left( \boldsymbol{r}_{0}\right)$. To investigate the specific scope of the far-field region, we compare in Fig.~\ref{modelv2} the optical binding forces corresponding to the exact calculation and the far-field approximation. As shown in Fig.~\ref{modelv2}, we can find that the forces experience oscillations, and that as the scaled displacement increases, the oscillation amplitudes of the optical binding forces decrease. This oscillation behavior can be understood because the forces are functions of the trigonometric functions, as shown in Eqs.~(\ref{31}). Meanwhile, it can be seen from Fig.~\ref{modelv2} that the exact optical-binding force is very close to the approximate optical-binding force when $R_{0}/\lambda>1$, which is consistent with the fact that the near-field constant is much smaller than the far-field constant when $R_{0}\sim\lambda$.

Since these interaction terms described by Eqs.~(\ref{crossterm}) involve two particles, below we analyze these cross interactions between the two particles together in the far-field regime.
The first cross term $\hat{H}_{\text{tw-Gtw}}=\hat{H}_{\text{tw-Gtw}}^{(1)}+\hat{H}_{\text{tw-Gtw}}^{(2)}$ describes the lateral binding of the two
identical spherical nanoparticles. It is given by%
\begin{eqnarray}
\hat{H}_{\text{tw-Gtw}} &\approx &-\frac{1}{2}\alpha\eta _{f_{\text{tw}}}( D/\hat{R}_{0})
E_{10}( \boldsymbol{\hat{R}}_{1}) E_{20}( \boldsymbol{\hat{R}}%
_{2})   \notag \\
&&\times \left\{ \cos \left[ k_{\text{tw}}(\hat{Z}_{0}+\hat{R}_{0})\right]\boldsymbol{e}_{\text{tw}}^{(1)}%
\cdot\overleftrightarrow{\mathbf{M}}_{f}( \boldsymbol{\hat{R}}_{0})\cdot
\boldsymbol{e}_{\text{tw}}^{(2)}\right. \notag \\
&&+ \left.\cos \left[ k_{\text{tw}}(\hat{Z}_{0}-\hat{R}_{0})\right]\boldsymbol{e}_{\text{tw}}^{(2)}\cdot \overleftrightarrow{\mathbf{M}}_{f}( \boldsymbol{\hat{R}}_{0})\cdot
\boldsymbol{e}_{\text{tw}}^{(1)}\right\} \text{,} \label{HTGT1}
\end{eqnarray}%
where $\eta_{f_{\text{tw}}}$ corresponds to the far-field constant for the wave number $k_{\text{tw}}$. The detailed derivation can be found in the Appendix. Expanding the corresponding electric field near the foci of tweezers 1
and 2, the $\hat{H}_{\text{tw-Gtw}}$ can be rewritten as
\begin{eqnarray}
\hat{H}_{\text{tw-Gtw}} &\approx & \hbar R_{\alpha} \left(
\hat{x}_{1}-\hat{x}_{2}\right)  \notag \\
&&+\sum_{q=x,y,z}\left[\nu_{q}(\hat{q}_{1}^{2}+\hat{q}_{2}^{2})+\frac{1}{2}k_{q}\left( \hat{q}_{1}-\hat{q}_{2}\right) ^{2}\right] , \label{18}
\end{eqnarray}%
where the first term corresponds to a shift of the equilibrium position of the center-of-mass motion along $x$-axis, and the displacement factor is given by
\begin{eqnarray}
R_{\alpha}&=&\alpha \eta _{f_{\text{tw}}}\epsilon _{\text{%
tw}}^{(1)}\epsilon _{\text{tw}}^{(2)}[ k_{\text{tw}}\sin \left( k_{\text{tw}%
}D\right) +D^{-1}\cos \left( k_{\text{tw}}D\right)]/\hbar \text{. }
\end{eqnarray}%
The second term in Eq.~(\ref{18}) describes the frequency shifts of the center-of-mass motion, with the frequency shifts
\begin{subequations}
\begin{align}
\nu_{x}&=\nu_{y}=\alpha \eta _{f_{\text{tw}}}\epsilon _{\text{tw}}^{(1)}\epsilon _{%
\text{tw}}^{(2)}\cos \left( k_{\text{tw}}D\right)/W_{t}^{2} ,   \\
\nu_{z}&=\alpha \eta _{f_{\text{tw}}}\epsilon _{\text{tw}}^{(1)}\epsilon _{%
\text{tw}}^{(2)}\cos \left( k_{\text{tw}}D\right)/(2z_{R}^{2}) .
\end{align}
\end{subequations}%
The third term in Eq.~(\ref{18}) describes the interaction between the two particles mediated by the light scattering, and the particle-particle coupling strengths are given by%
\begin{subequations}
\begin{align}
k_{x}=&-\alpha \eta _{f_{\text{tw}}}\epsilon _{\text{tw}}^{(1)}\epsilon _{%
\text{tw}}^{(2)}[ ( 2D^{-2}-k_{\text{tw}}^{2}) \cos \left( k_{%
\text{tw}}D\right)    \notag \\
& +2k_{\text{tw}}D^{-1}\sin \left( k_{\text{tw}}D\right) ] , \label{kx}\\
k_{y} =&\alpha \eta _{f_{\text{tw}}}\epsilon _{\text{tw}}^{(1)}\epsilon _{\text{%
tw}}^{(2)}[ 3D^{-2}\cos \left( k_{\text{tw}}D\right)    \notag \\
& +k_{\text{tw}}D^{-1}\sin \left( k_{\text{tw}}D\right) ] ,
\\
k_{z} =&\alpha \eta _{f_{\text{tw}}}\epsilon _{\text{tw}}^{(1)}\epsilon _{\text{%
tw}}^{(2)}[ ( D^{-2}+k_{\text{tw}}^{2}) \cos \left( k_{\text{tw}%
}D\right)    \notag \\
& +k_{\text{tw}}D^{-1}\sin \left( k_{\text{tw}}D\right) ] .
\end{align}
\end{subequations}%

The second cross term $\hat{H}_{\text{cav-Gcav}}$ given by Eq.~(\ref{13}) represents the longitudinal binding of the two spherical
nanoparticles,
\begin{eqnarray}
\hat{H}_{\text{cav-Gcav}} &=&\hat{H}_{\text{cav-Gcav}}^{(1)}+\hat{H}_{\text{cav-Gcav}}^{(2)} \notag \\
&\approx &-4\alpha\eta
_{f} \epsilon _{\text{cav}}^{2}( D/\hat{R}_{0}) \cos ( k\hat{X}_{1}) \cos (
k\hat{X}_{2})   \notag \\
&&\times \cos ( k\hat{R}_{0}) ( \hat{a}^{\dagger }\hat{a}+1/2)
\boldsymbol{e}_{\text{cav}}\cdot\overleftrightarrow{\mathbf{M}}_{f}( \boldsymbol{\hat{R}}_{0})\cdot
\boldsymbol{e}_{\text{cav}}, \label{HCGC1}
\end{eqnarray}%
where $\eta_{f}$ corresponds to the far-field constant for the wave number $k$. The Hamiltonian $\hat{H}_{\text{cav-Gcav}}$ can be further re-expressed as%
\begin{eqnarray}
\hat{H}_{\text{cav-Gcav}} &\approx &-4\alpha\eta
_{f} \epsilon _{\text{cav}}^{2}\cos( kx_{10}) \cos( kx_{20}) \cos ( kD) \hat{a}^{\dagger }\hat{a}
\notag \\
&&+\hbar g_{\alpha}\hat{x}_{1}( \hat{a}^{\dagger }\hat{a}+1/2)-\hbar g_{\alpha
}\hat{x}_{2}( \hat{a}^{\dagger }\hat{a}+1/2), \label{HCGC}
\end{eqnarray}%
where the first term is the frequency shift term and the last two terms
are the optomechanical coupling terms, with the coupling strength%
\begin{eqnarray}
g_{\alpha } &=&4\alpha\eta _{f} \epsilon _{\text{cav}}^{2}[ (
k\sin \left( kD\right) +D^{-1}\cos \left( kD\right) )
\notag \\
&& \times \cos ^{2}\left( kD/2\right) +k\sin \left(
kD\right) \cos \left( kD\right)/2 ]/\hbar.
\end{eqnarray}

The third cross term $\hat{H}_{\text{tw-Gcav}}$ describes the interaction between the $j$th tweezer
field at $\boldsymbol{\hat{R}}_{j}$ and the field generated by the
other dipole caused by the cavity field. This term reads
\begin{eqnarray}
\hat{H}_{\text{tw-Gcav}} &=&\hat{H}_{\text{tw-Gcav}}^{(1)}+\hat{H}_{\text{tw-Gcav}}^{(2)} \notag \\
&\approx &-\frac{1}{2}\alpha\eta
_{f} \epsilon _{\text{cav}}( D/\hat{R}_{0}) \cos ( k\hat{R}_{0}) E_{10}(
\boldsymbol{\hat{R}}_{1}) \cos ( k\hat{X}_{2})   \notag \\
&&\times \left( \hat{a}^{\dagger }e^{-ik_{\text{tw}}\hat{Z}_{1}}+\hat{a}e^{ik_{\text{tw}%
}\hat{Z}_{1}}\right) \boldsymbol{e}_{\text{tw}}^{(1)}\cdot\overleftrightarrow{\mathbf{M}}_{f}( \boldsymbol{\hat{R}%
}_{0}) \cdot\boldsymbol{e}_{\text{cav}}  \notag \\
&&-\frac{1}{2}\alpha \epsilon _{\text{cav}}\eta _{f}( D/\hat{R}_{0})%
 \cos ( k\hat{R}_{0}) E_{20}( \boldsymbol{\hat{R}}_{2})
\cos ( k\hat{X}_{1})   \notag \\
&&\times \left( \hat{a}^{\dagger }e^{-ik_{\text{tw}}\hat{Z}_{2}}+\hat{a}e^{ik_{\text{tw}%
}\hat{Z}_{2}}\right) \boldsymbol{e}_{\text{tw}}^{(2)}\cdot\overleftrightarrow{\mathbf{M}}_{f}( \boldsymbol{\hat{R}%
}_{0})\cdot \boldsymbol{e}_{\text{cav}}, \label{HTGC1}
\end{eqnarray}%
which can be rewritten as%
\begin{eqnarray}
\hat{H}_{\text{tw-Gcav}} &\approx & \hbar \Omega_{\alpha}( \hat{a}^{\dagger }+\hat{a})
+\sum_{j=1,2}\hbar g_{\alpha x_{j}}( \hat{a}^{\dagger }+\hat{a}) \hat{x}_{j}  \notag \\
&&+\sum_{j=1,2}i\hbar g_{\alpha z_{j}}(\hat{a}-\hat{a}^{\dagger }) \hat{z}_{j}. \label{19}
\end{eqnarray}%
In Eq.~(\ref{19}), we introduced $\Omega_{\alpha}=-\alpha\eta
_{f} \epsilon _{\text{cav}}\cos \left( kD\right)\cos \left( kD/2\right) ( \epsilon _{\text{tw}}^{(1)}+\epsilon _{\text{%
tw}}^{(2)})/(2\hbar)$ and the coupling strengths%
\begin{subequations}
\begin{align}
g_{\alpha x_{1}} =&\frac{1}{2\hbar}\alpha \eta _{f}\epsilon _{\text{cav}}[
( k\sin \left( kD\right) +D^{-1}\cos \left( kD\right) )
( \epsilon _{\text{tw}}^{(1)}+\epsilon _{\text{tw}}^{(2)})
\notag \\
&\times \left. \cos \left( kD/2\right) +k\cos \left( kD\right)
\epsilon _{\text{tw}}^{(2)}\sin \left( kD/2\right) \right],  \\
g_{\alpha x_{2}} =&-\frac{1}{2\hbar}\alpha \eta _{f}\epsilon _{\text{cav}}[
( k\sin \left( kD\right) +D^{-1}\cos \left( kD\right) )
( \epsilon _{\text{tw}}^{(1)}+\epsilon _{\text{tw}}^{(2)})
\notag \\
&\times \left. \cos \left( kD/2\right) +k\cos \left( kD\right)
\epsilon _{\text{tw}}^{(1)}\sin \left( kD/2\right) \right],  \\
g_{\alpha z_{j}} =&\frac{1}{2\hbar}\alpha \eta _{f}\epsilon _{\text{cav}%
}\epsilon _{\text{tw}}^{(j)}k_{\text{tw}}\cos \left( kD\right) \cos \left( k%
D/2\right).
\end{align}
\end{subequations}%

Finally, the interaction Hamiltonian between the cavity field and the field emitted by the
dipole induced by the tweezer fields reads
\begin{eqnarray}
\hat{H}_{\text{cav-Gtw}} &=& \hat{H}_{\text{cav-Gtw}}^{(1)}+\hat{H}_{\text{cav-Gtw}}^{(2)} \notag \\
&\approx &-\frac{1}{2}\alpha \eta
_{f_{\text{tw}}}\epsilon _{\text{cav}}( D/\hat{R}_{0}) \cos ( k\hat{X}_{1})
E_{20}( \boldsymbol{\hat{R}}_{2}) \left( \hat{a}e^{-ik_{\text{tw}}(
\hat{R}_{0}-\hat{Z}_{2}) }\right.   \notag \\
&&\left. +\hat{a}^{\dagger }e^{ik_{\text{tw}}( \hat{R}_{0}-\hat{Z}_{2}) }\right)
\boldsymbol{e}_{\text{cav}}\cdot\overleftrightarrow{\mathbf{M}}_{f}( \boldsymbol{\hat{R}}%
_{0}) \cdot\boldsymbol{e}_{\text{tw}}^{(2)}  \notag \\
&&-\frac{1}{2}\alpha \epsilon _{\text{cav}}\eta _{f_{\text{tw}}}( D/\hat{R}_{0})%
 \cos ( k\hat{X}_{2}) E_{10}( \boldsymbol{\hat{R}}%
_{1}) \left( \hat{a}e^{-ik_{\text{tw}}( \hat{R}_{0}-\hat{Z}_{1}) }\right.
\notag \\
&&\left. +\hat{a}^{\dagger }e^{ik_{\text{tw}}( \hat{R}_{0}-\hat{Z}_{1}) }\right)
\boldsymbol{e}_{\text{cav}}\cdot\overleftrightarrow{\mathbf{M}}_{f}( \boldsymbol{\hat{R}}%
_{0}) \cdot\boldsymbol{e}_{\text{tw}}^{(1)} , \label{HCGT1}
\end{eqnarray}%
which can be further expressed as%
\begin{eqnarray}
\hat{H}_{\text{cav-Gtw}} &\approx &
 \hbar(\Omega_{\beta}\hat{a}+\Omega_{\beta}^{*} \hat{a}^{\dagger })
+\sum_{j=1,2}\hbar( g_{\beta x_{j}}\hat{a}+g_{\beta x_{j}}^{\ast }\hat{a}^{\dagger
}) \hat{x}_{j}  \notag \\
&&+\sum_{j=1,2}i\hbar( g_{\beta z_{j}}\hat{a}-g_{\beta z_{j}}^{\ast }\hat{a}^{\dagger
}) \hat{z}_{j} \text{.} \label{HCGT}
\end{eqnarray}%
Here, the displacement factor of mode $a$ is $\Omega_{\beta}=-\alpha\eta _{f_{%
\text{tw}}} \epsilon _{\text{cav}%
}( \epsilon _{\text{tw}}^{(1)}+\epsilon _{\text{tw}}^{(2)})\cos \left( kD/2\right)e^{-ik_{\text{tw}}D}/(2\hbar)$ and the coupling strengths are
\begin{subequations}
\begin{align}
g_{\beta x_{1}} =&\frac{1}{2\hbar}\alpha \eta _{f_{\text{tw}}}\epsilon _{\text{%
cav}}[ (D^{-1}+ik_{\text{tw}}) ( \epsilon _{\text{%
tw}}^{(1)}+\epsilon _{\text{tw}}^{(2)})    \notag \\
&\left. \times \cos \left( kD/2\right) +\epsilon _{\text{tw}%
}^{(2)}k\sin \left( kD/2\right) \right] e^{-ik_{\text{tw}}D}, \\
g_{\beta x_{2}} =&-\frac{1}{2\hbar}\alpha \eta _{f_{\text{tw}}}\epsilon _{\text{%
cav}}[ (D^{-1}+ik_{\text{tw}}) ( \epsilon _{\text{%
tw}}^{(1)}+\epsilon _{\text{tw}}^{(2)})    \notag \\
&\left. \times \cos \left( kD/2\right) +\epsilon _{\text{tw}%
}^{(1)}k\sin \left( kD/2\right) \right] e^{-ik_{\text{tw}}D}, \\
g_{\beta z_{j}} =&-\frac{1}{2\hbar}\alpha \eta _{f_{\text{tw}}}\epsilon _{\text{%
cav}}\epsilon _{\text{tw}}^{(j)}k_{\text{tw}}\cos \left( kD/2\right)
e^{-ik_{\text{tw}}D}.
\end{align}
\end{subequations}%

Based on the above analyses, we obtain the total Hamiltonian in the rotating frame [defined by the unitary transformation operator $\exp(-i\omega_{\text{tw}}\hat{a}^{\dagger}\hat{a}t)$] as
\begin{eqnarray}
\hat{H}_{\text{tot}} &=&\hbar \Delta ^{\prime }\hat{a}^{\dagger }\hat{a}+\sum_{j=1,2}\frac{%
\hat{\mathbf{P}}_{j}^{2}}{2m}+\sum_{j=1,2}\sum_{Q=x,y,z}\frac{1}{2}m_{j}\tilde{%
\omega}_{jQ}^{2}\hat{Q}_{j}^{2}  \notag \\
&&+\sum_{j=1,2}\hbar \tilde{g}_{ax_{j}}\hat{a}^{\dagger }\hat{a}\hat{x}_{j}-\sum_{Q=x,y,z}
k_{Q}\hat{Q}_{1}\hat{Q}_{2}  \notag \\
&&+\hbar ( \tilde{\Omega} \hat{a}+\tilde{\Omega} ^{\ast }\hat{a}^{\dagger }) +\hbar \tilde{R}(
\hat{x}_{1}-\hat{x}_{2})   \notag \\
&&+\hbar \sum_{j=1,2}[ \hat{a}( \tilde{g}_{x_{j}}\hat{x}_{j}+i\tilde{g}_{z_{j}}\hat{z}_{j}) +%
\text{H.c.}] , \label{Htot}
\end{eqnarray}
where the effective driving detuning is $\Delta ^{\prime }=\Delta -2\alpha \epsilon _{\text{cav}}^{2}\cos
^{2}\left( kD/2\right)/\hbar -4\alpha \epsilon _{\text{cav}}^{2}\eta _{f}\cos
^{2}\left( kD/2\right) \cos \left( kD\right)/\hbar $ with $\Delta=\omega_{\text{cav}}-\omega_{\text{tw}}$. The $j$th particle exhibits a $Q$-mode frequency of  $\tilde{\omega}_{jQ}=\sqrt{\omega_{jQ}^{2}+2\nu_{Q}/m+k_{Q}/m}$ with $j=1,2$ and $Q=x,y,z$,  the displacement factor of the cavity mode and $x$ modes are, respectively, given by $\tilde{\Omega}=\Omega+\Omega_{\alpha}+\Omega_{\beta}$ and $\tilde{R}_{j}=R_{\alpha}+g_{\alpha}/2$, and the optomechanical couplings are given by $\tilde{g}_{ax_{j}}=g_{ax_{j}}-(-1)^{j}g_{\alpha }$ and $\tilde{g}_{x(z)_{j}}=g_{x(z)_{j}}+g_{\alpha x(z)_{j}}+g_{\beta x(z)_{j}}$. It can be seen from Eq.~(\ref{Htot}) that the $y$ modes of the two
particles are only coupled to each other and decoupled from the other modes, so we
will only consider the $x$ and $z$ modes in the following discussions. It should be pointed out that if the polarization direction of the two optical tweezers is not perpendicular to the cavity axis, the coupling channel between the $y$ modes and the other modes will be turned on. We also note that, to be concise, the hat for all the operators will be ignored in the following.

\section{Simultaneous ground-state cooling of the x-direction motions}\label{Cooling to two modes ground-state}
For cooling of the $x$-direction (along the cavity axis) motion of a single-levitated nanoparticle,
we prefer to locate the particle at the nodes of the cavity mode $|\sin\left(kx_{0}\right)|=1$~\cite{UPRL2019}.
Below we consider that the two particles are located at $x_{10}=D/2$ and $x_{20}=-D/2$,
which satisfy $\sin (kx_{10})=1$ and $\sin (kx_{20})=-1$. To cool the mechanical
modes, we consider the red-sideband resonance regime: $\Delta =\omega _{\text{cav}}-\omega
_{\text{tw}}=\omega _{\text{m}}$. Concretely, we assume that the tweezer laser has the wavelength $\lambda _{\text{tw}}=1064$ nm and the trapping
frequency of the particles is $\omega _{\text{m}}/2\pi \sim 100$
MHz. Therefore, the driving frequency is much larger than the resonance
frequency of the oscillator $\omega _{\text{tw}}\gg \omega _{\text{m}}$, then the
wave number of the cavity field is approximately equal to that of the
tweezer field $k\approx k_{\text{tw}}$, and we can make the following
approximations: $\sin \left( k_{\text{tw}}D/2\right) \approx 1$, $\cos
\left( k_{\text{tw}}D/2\right) \approx 0$, $\cos \left( kD\right) \approx
\cos (k_{\text{tw}}D)\approx -1$, $\sin \left( kD\right) \approx \cos (k_{%
\text{tw}}D)\approx 0$, and $e^{-ik_{\text{tw}}D}\approx e^{ik_{\text{tw}%
}D}\approx -1$. In this case, the $z$ modes of the two particles are decoupled from other modes ($x$ modes and the cavity mode $a$). The $z$ modes of the two particles are coupled to each other via the $z$-$z$ coupling, and the $z$-mode motion cannot be cooled because they are connected with the environments and decoupled from the cooling channel. For studying the cooling of mechanical motion, in this section we focus on the $x$ modes and cavity mode. In this case, the Hamiltonian of the system including the $x$ modes and cavity mode is reduced to%
\begin{eqnarray}
H_{\text{tot}} &=&\hbar \Delta a^{\dagger }a+\sum_{j=1,2}\left( \frac{%
P_{jx}^{2}}{2m}+\frac{m\Omega _{j}^{2}x_{j}^{2}}{2}\right)+\hbar R\left(
x_{1}-x_{2}\right)  \notag \\
&&+\sum_{j=1,2}\hbar g_{j}( a^{\dagger }+a) x_{j} -k_{x}x_{1}x_{2}, \label{21}
\end{eqnarray}%
where the $x$-mode oscillation frequency of the $j$th nanoparticle is given by $\Omega _{j}^{2}=( %
\alpha \epsilon _{\text{tw}}^{(j)^{2}}/{W_{t}^{2}}-2\alpha \eta _{f_{\text{%
tw}}}\epsilon _{\text{tw}}^{(1)}\epsilon _{\text{tw}}^{(2)}/{W_{t}^{2}}%
+k_{x})/m $, the coupling strengths are given by%
\begin{subequations}
\begin{align}
g_{1} &=\alpha \epsilon _{\text{cav}}[ \epsilon _{%
\text{tw}}^{(1)}-( \eta _{f}+\eta _{f_{\text{tw}}}) \epsilon _{\text{%
tw}}^{(2)}] k/(2\hbar), \\
g_{2} &=-\alpha \epsilon _{\text{cav}}[ \epsilon _{%
\text{tw}}^{(2)}-( \eta _{f}+\eta _{f_{\text{tw}}}) \epsilon _{\text{%
tw}}^{(1)}] k/(2\hbar),
\end{align}
\end{subequations}%
and the displacement factor is
\begin{equation}
R=-\alpha \eta _{f_{\text{tw}}}\epsilon _{\text{tw}}^{(1)}\epsilon _{\text{tw}}^{(2)}/(\hbar D).%
\end{equation}%
We see from Hamiltonian~(\ref{21}) that there exist bilinear couplings between the
cavity field and the $x$-direction motions. In addition, the two harmonic oscillations are coupled with each other via the $x$-$x$ interaction, and both the two oscillators are displaced in the $x$-direction. For analyzing the cooling of the $x$-direction motion, below we will
work in the displaced representation of the system such that the excitations are associated with the fluctuations.
For convenience, we introduce the dimensionless position and momentum operators $\sqrt{2}q_{j=1,2}=x_{j}/x_{j\text{,zpf}}$ and $\sqrt{2}p_{j=1,2}=P_{jx}/p_{j\text{,zpf}}%
$, where the zero-point motions are $x_{j\text{,zpf}}=\sqrt{ \hbar/(2m\Omega_{j}) }$ and $p_{j\text{,zpf}}=\sqrt{ m\Omega_{j}\hbar/2 }$. We also introduce the quadrature operators $X=(a^{\dagger }+a) /\sqrt{2}$ and $Y=i( a^{\dagger }-a) /\sqrt{%
2}$ for the cavity field.

Based on Eq.~(\ref{21}), we can obtain the quantum Langevin equations for the system as%
\begin{subequations}
\label{22}
\begin{align}
\dot{q}_{1} =&\Omega _{1}p_{1},  \label{6} \\
\dot{p}_{1} =&-\Omega _{1}q_{1}-\sqrt{2}G_{1}X-R_{1}+G_{x}q_{2}-\gamma
_{1}p_{1}+f_{\text{th}}^{(1)},   \label{7} \\
\dot{q}_{2} =&\Omega _{2}p_{2},  \label{8} \\
\dot{p}_{2} =&-\Omega _{2}q_{2}-\sqrt{2}G_{2}X+R_{2}+G_{x}q_{1}-\gamma
_{2}p_{2}+f_{\text{th}}^{(2)},   \label{9} \\
\dot{X} =&\Delta Y-\kappa X+\sqrt{2\kappa }X_{\text{in}},
\label{10} \\
\dot{Y} =&-\Delta X-\kappa Y-\sqrt{2}\sum_{j=1,2}G_{j}q_{j}+\sqrt{2\kappa }Y_{%
\text{in}}\text{,}   \label{11}
\end{align}
\end{subequations}%
where $\kappa$ and $\gamma_{j}$ are the decay rates of the cavity mode and the $x$-direction motion of the $j$th particle,
respectively. In Eqs.~(\ref{22}), we introduce $G_{j}=\sqrt{2}g_{j}x_{j,\text{zpf}}$, $G_{x}=2k_{x}x_{1,\text{zpf}}x_{2,\text{zpf}}/\hbar $, and $%
R_{j}=\sqrt{2}Rx_{j,\text{zpf}}$. The $f_{\text{th}}^{(j)}$ is the stochastic thermal noise
operator corresponding to the $x$-mode motion of the $j$th particle, which is determined by
the zero average values
\begin{equation}
\left\langle f_{\text{th}}^{(j)}\left( t\right)\right\rangle =0,\hspace{0.5cm}j=1,2,
\end{equation}%
and the correlation function,
\begin{equation}
\left\langle f_{\text{th}}^{(j)}\left( t\right) f_{\text{th}}^{(j^{\prime })}\left( t^{\prime
}\right) \right\rangle =\delta _{jj^{\prime }}\frac{\gamma _{j}}{\Omega _{j}}\int e^{-i\omega
\left( t-t^{\prime }\right) }\omega \left[ \coth \left( \frac{\hbar \omega }{%
2k_{B}T_{j}}\right) +1\right] \frac{d\omega }{2\pi }, \label{27}
\end{equation}%
where $k_{B}$ is the Boltzmann constant, and $T_{j}$ is the temperature of the thermal
bath associated with the $x_{j}$ mode. We assume that these mechanical modes
are connected to the high-temperature reservoirs ($k_{B}T_{j}\gg\hbar\Omega_{j}$),
so we can obtain $\coth [ \hbar \Omega_{j}/ ({k_{B}T_{j}})] +1\approx 2k_{B}T_{j}/%
(\hbar \Omega_{j}) $ in the high-temperature limit. The stochastic noise is
reduced to a delta-correlation noise $\langle f_{\text{th}}^{(j)}\left( t\right)
f_{\text{th}}^{(j^{\prime })}\left( t^{\prime }\right) +f_{\text{th}}^{(j^{\prime })}\left(
t^{\prime }\right) f_{\text{th}}^{(j)}\left( t\right) \rangle \approx 2\gamma
_{j}( 2\bar{n}_{j\text{,th}}+1) \delta \left( t-t^{\prime }\right) \delta
_{jj^{\prime }}$, where $\bar{n}_{j,\text{th}}=[ \exp [ \hbar \Omega_{j} /(k_{B}T_{j})%
]-1 ] ^{-1}\approx k_{B}T_{j}/( \hbar \Omega_{j} ) $ is the
thermal occupation number for the $j$th thermal bath. In addition, the $X_{\text{in}}=( a_{%
\text{in}}^{\dagger }+a_{\text{in}}) /\sqrt{2}$ and $Y_{\text{in }%
}=i( a_{\text{in}}^{\dagger }-a_{\text{in}}) /\sqrt{2}$ are the optical noise operators, which
are determined by the zero average values $\langle X_{\text{in}}\rangle=0$ and $\langle Y_{\text{in}} \rangle=0$. The correlation functions of these optical noise operators are given by~\cite{Cbook2000}%
\begin{subequations}
\label{28}
\begin{align}
\left\langle  X_{\text{in}}\left( t\right)  X_{\text{in}}\left(
t^{\prime }\right) \right\rangle  &=\frac{1%
}{2}\delta \left( t-t^{\prime }\right) , \\
\left\langle  Y_{\text{in}}\left(
t\right)  Y_{\text{in}}\left( t^{\prime }\right) \right\rangle &=\frac{1%
}{2}\delta \left( t-t^{\prime }\right) , \\
\langle  X_{\text{in}}\left( t\right)  Y_{\text{in}}\left(
t^{\prime }\right) \rangle  &=\frac{i}{2}\delta \left( t-t^{\prime }\right)
, \\
\langle  Y_{\text{in}}\left( t\right)  X_{\text{in}}\left(
t^{\prime }\right) \rangle  &=-\frac{i}{2}\delta \left( t-t^{\prime
}\right) .
\end{align}
\end{subequations}%
\begin{figure}[!t]
\center\includegraphics[width=0.48\textwidth]{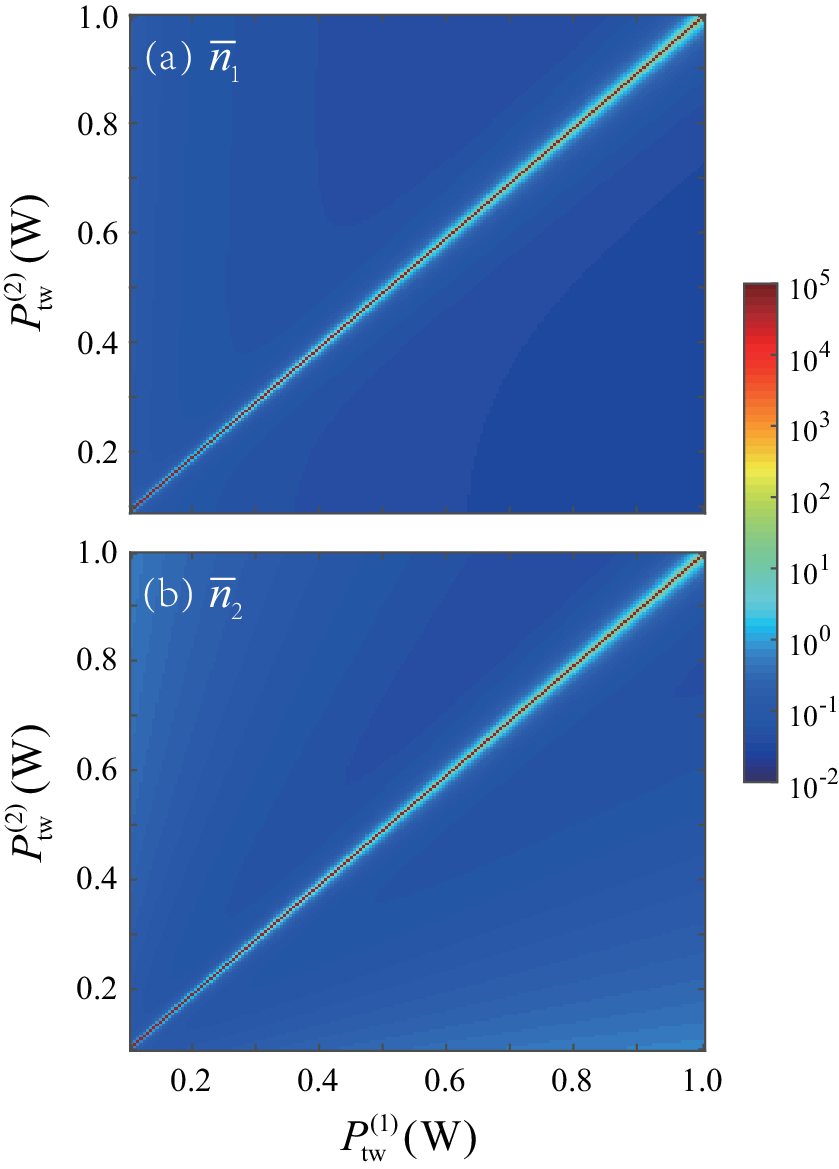}
\caption{{ The final mean phonon numbers (a) $\bar{n}_{1}$ and (b) $\bar{n}_{2}$  in the two mechanical modes versus the powers $P_{%
\text{tw}}^{(1)}$ and $P_{\text{tw}}^{(2)}$ of the two tweezers. The
silica nanoparticle of radius is $a_{0}\approx 90$ nm, the separation of
the particles is $D\approx 2.5\lambda $, the initial occupations are $%
\bar{n}_{1\text{,th}}=\bar{n}_{2\text{,th}}= 10^{5}$, and the effective driving detuning is $\Delta =
\Omega _{1}$. Other parameters used are $\gamma _{1}/\Omega_{1}=\gamma
_{2}/\Omega _{1}= 0.5\times10^{-8}$ and $\kappa/\Omega_{1}= 0.2.$}}
\label{modelv3}
\end{figure}%

To work in the displacement representation, we re-express Eqs.~(\ref{6})--(\ref{11}) around the steady-state values by writing
operators $O\in \{q_{j=1,2},p_{j=1,2},X,Y\} $ as the sum of average value
and quantum fluctuation: $O=\left\langle O\right\rangle +\delta O$%
. Then, the Langevin equations can be separated into the semi-classical equations of
motion and the equations of motion for quantum fluctuations. The latter can be written as
\begin{subequations}
\label{23}
\begin{align}
\delta \dot{q}_{1} &=\Omega _{1}\delta p_{1}, \\
\delta \dot{p}_{1} &=-\Omega _{1}\delta q_{1}-\sqrt{2}G_{1}\delta
X+G_{x}\delta q_{2}-\gamma _{1}\delta p_{1}+f_{\text{th}}^{(1)},  \\
\delta \dot{q}_{2} &=\Omega _{2}\delta p_{2}, \\
\delta \dot{p}_{2} &=-\Omega _{2}\delta q_{2}-\sqrt{2}G_{2}\delta
X+G_{x}\delta q_{1}-\gamma _{2}\delta p_{2}+f_{\text{th}}^{(2)},  \\
\delta \dot{X} &=\Delta \delta Y-\kappa \delta X+\sqrt{2\kappa }\delta X_{%
\text{in}},  \\
\delta \dot{Y} &=-\Delta \delta X-\kappa \delta Y-\sum_{j=1,2}\sqrt{2}%
G_{j}\delta q_{j}+\sqrt{2\kappa }\delta Y_{\text{in}}.
\end{align}%
\end{subequations}%
By introducing the operator vector
\begin{equation}
\mathbf{u}\left( t\right)=(\delta q_{1},\delta p_{1}, \delta q_{2},  \delta p_{2} , \delta X , \delta Y) ^{\text{T}},
\end{equation}%
with ``T'' denotes the matrix transpose, and the noise operator vector
\begin{equation}
\mathbf{N}\left( t\right) =(
0 , f_{\text{th}}^{(1)}\left( t\right) , 0 , f_{\text{th}}^{(2)}\left( t\right) , \sqrt{%
2\kappa }\delta X_{\text{in}}\left( t\right) , \sqrt{2\kappa }\delta Y_{%
\text{in}}\left( t\right) ) ^{\text{T}},
\end{equation}%
Eqs.~(\ref{23}) can be expressed as a compact matrix form
\begin{equation}
\mathbf{\dot{u}}\left( t\right) =\mathbf{Au}\left( t\right) +\mathbf{N}%
\left( t\right),
\end{equation}%
where the coefficient matrix $\mathbf{A}$ is given by
\begin{equation}
\mathbf{A}=\left(
\begin{array}{cccccc}
0 & \Omega _{1} & 0 & 0 & 0 & 0 \\
-\Omega _{1} & -\gamma _{1} & G_{x} & 0 & -\sqrt{2}G_{1} & 0 \\
0 & 0 & 0 & \Omega _{2} & 0 & 0 \\
G_{x} & 0 & -\Omega _{2} & -\gamma _{2} & -\sqrt{2}G_{2} & 0 \\
0 & 0 & 0 & 0 & -\kappa & \Delta \\
-\sqrt{2}G_{1} & 0 & -\sqrt{2}G_{2} & 0 & -\Delta & -\kappa%
\end{array}%
\right).
\end{equation}%

To calculate the final mean phonon numbers in these mechanical modes, we introduce the covariance matrix
$\mathbf{V}$ defined by the matrix elements
\begin{equation}
\mathbf{V}_{nm}=\frac{1}{2}[\mathbf{u}_{n}(\infty)\mathbf{u}_{m}(\infty)+\mathbf{u}_{m}(\infty)\mathbf{u}_{n}(\infty)]
\end{equation}%
for $n,m=1$-$6$. The covariance matrix $\mathbf{V}$ is determined by the Lyapunov equation~\cite{DPRL2007}%
\begin{equation}
\mathbf{AV}+\mathbf{VA}^{\text{T}}=-\mathbf{Q},
\end{equation}%
where $\mathbf{Q}$ is the noise correlation matrix, defined by the elements $\mathbf{Q}_{nm}=\frac{1}{2}\langle
\mathbf{N}_{n}(t)\mathbf{N}_{m}(t^{\prime })+\mathbf{N}_{m}(t)\mathbf{N}_{n}(t^{\prime })\rangle $ for $n,m=1$-$6$.
Based on Eqs.~(\ref{27}) and (\ref{28}), the noise correlation matrix can be obtained as
\begin{equation}
\mathbf{Q}=\text{diag}\left[
0,\gamma_{1}\left( 2\bar{n}_{1,\text{th}}+1\right),0,\gamma_{2}\left( 2\bar{n}_{2,\text{th}}+1\right),\kappa,\kappa\right].
\end{equation}%
The final mean phonon numbers of the two mechanical modes can be
expressed as%
\begin{equation}
 \bar{n}_{j} =\frac{1}{2}\left[ \langle \delta
q_{j}^{2}\rangle +\langle \delta p_{j}^{2}\rangle -1\right],\hspace{0.5cm}j=1,2,
\end{equation}%
where the stationary variance of the mechanical modes is given by the
corresponding diagonal matrix elements of the covariance matrix,
\begin{subequations}
\begin{align}
\langle \delta q_{1}^{2}\rangle &=\mathbf{V}_{11},\hspace{0.5cm} \langle \delta p_{1}^{2}\rangle=\mathbf{V}_{22},\\
\langle \delta q_{2}^{2}\rangle &=\mathbf{V}_{33},\hspace{0.5cm} \langle \delta p_{2}^{2}\rangle=\mathbf{V}_{44} \text{.}
\end{align}
\end{subequations}%
Therefore, the final mean phonon numbers in the two mechanical modes can be obtained by solving the Lyapunov equation. Note that the stability conditions for the steady states in our simulations have been confirmed with the Routh-Hurwitz criterion~\cite{Ibook2014}.

In the coupled cavity-levitated-nanoparticles system, the powers of the optical tweezers affect both the resonance frequencies of the mechanical modes and the coupling strengths between the cavity mode and the mechanical modes. Below, we analyze the dependence of the cooling efficiency of the $x$ modes of the two nanoparticles
on the powers of the two optical tweezers. In Fig.~\ref{modelv3}, we plot the final mean
phonon numbers $\bar{n}_{1}$ and $\bar{n}_{2}$ as functions of the powers $P_{\text{tw}}^{(1)}$
and $P_{\text{tw}}^{(2)}$. We can see from Fig.~\ref{modelv3} that the cooling of the two modes $x_{1}$ and $x_{2}$ is strongly suppressed when the two tweezers have the same power. The cooling performance becomes better when the working point deviates the diagonal line $P_{\text{tw}}^{(1)}=P_{\text{tw}}^{(2)}$. This phenomenon can be well
explained based on the dark-mode effect~\cite{JNP2023,CNJP2008,WangPRL2012,TianPRL2012,DPRA2020,DPRL2022,JPRA2022a,Jarx2023}.
From the expressions of $\Omega_{j}$ and $G_{j}$, we know that, when the powers of the two optical
tweezers are equal, i.e., $P_{\text{tw}}^{(1)}=P_{\text{tw}}^{(2)}$, the two modes $x_{1}$ and $x_{2}$ have
the same resonance frequency $\Omega_{1}=\Omega_{2}$, and the optomechanical coupling strengths are equal but with opposite signs $%
G_{1}=-G_{2}$.
In the following, we analyze the dark-mode effect in the system under these parameters.
\begin{figure}[!t]
\center\includegraphics[width=0.48\textwidth]{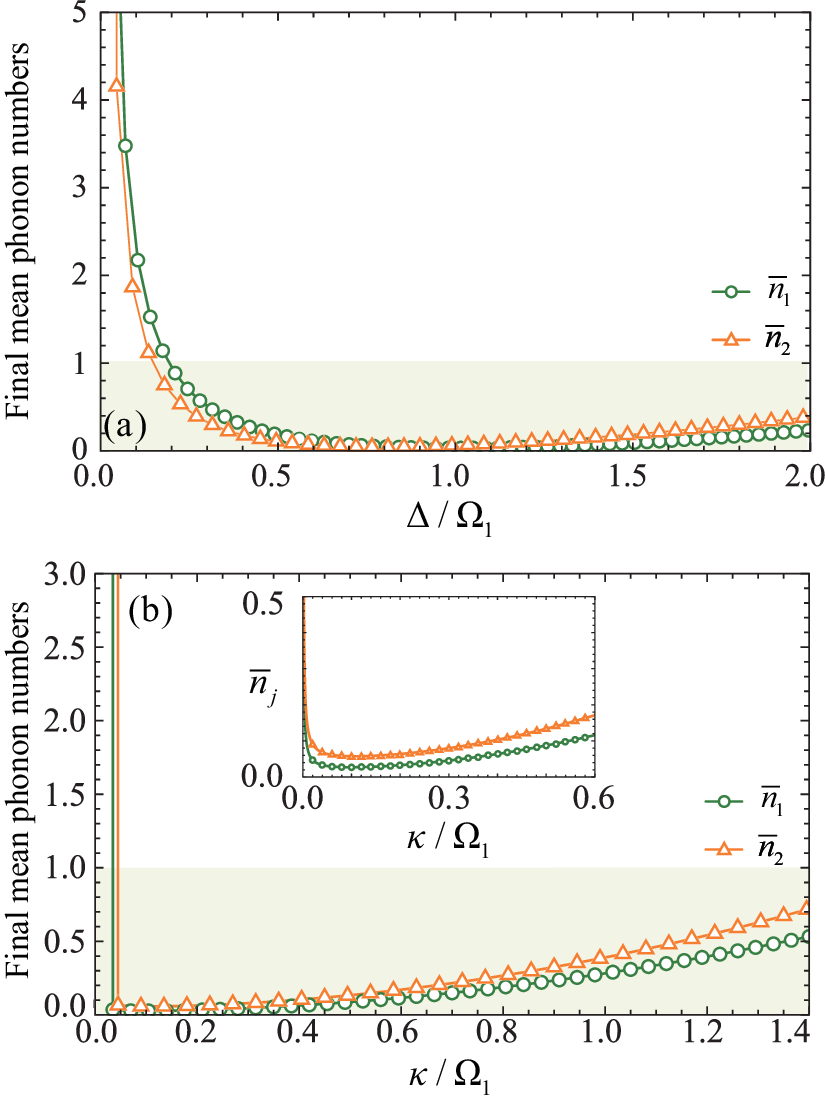}
\caption{{(a) The final mean phonon numbers $\bar{n}_{1}$ (green solid line with circles) and $\bar{n}_{2}$
(yellow solid line with triangles) versus the scaled
driving detuning $\Delta/\Omega_{1} $ when $\kappa/\Omega_{1}=0.2$. (b) The
final average phonon numbers $\bar{n}_{1}$ and $\bar{n}_{2}$ versus the scaled  cavity linewidth $\kappa/\Omega_{1} $ at $\Delta/\Omega_{1}=1$. The inset is a zoom-in plot of the phonon numbers in a narrower ranger of $\kappa/\Omega_{1}.$
Other common parameters used are $P_{\text{tw}}^{(1)}=0.8$~W, $P_{\text{tw}}^{(2)}=0.45$~W,  $\bar{n}_{\text{1,th}}=\bar{n}_{\text{2,th}}=
10^{5}$, $\Omega _{2}/\Omega _{1}\approx
0.75\text{, }G_{1}/\Omega _{1}\approx 0.22\text{, }G_{2}/\Omega _{1}\approx
-0.19\text{, }G_{x}/\Omega _{1}\approx -0.046\text{, and }\gamma _{1}/\Omega _{1}=\gamma
_{2}/\Omega _{1}= 0.5\times10^{-8}$. }}
\label{modelv4}
\end{figure}%

Using the dimensionless operators, the Hamiltonian characterizing the quantum fluctuations can be written as%
\begin{equation}
H_{\text{lin}}/\hbar =\Delta a^{\dagger }a+\sum_{j=1,2}\left[\frac{\Omega _{j}}{2}%
( p_{j}^{2}+q_{j}^{2}) +G_{j}( a^{\dagger }+a)
q_{j}\right]-G_{x}q_{1}q_{2}.  \label{16}
\end{equation}%
To clearly see the dark-mode effect in the system, we define two hybrid
modes of the two mechanical modes as%
\begin{subequations}
\label{24}
\begin{align}
q_{+}&=\frac{G_{1}q_{1}+G_{2}q_{2}}{\sqrt{G_{1}^{2}+G_{2}^{2}}}\text{, }\hspace{0.5cm} p_{+}=%
\frac{G_{1}p_{1}+G_{2}p_{2}}{\sqrt{G_{1}^{2}+G_{2}^{2}}}, \\
q_{-}&=\frac{G_{1}q_{2}-G_{2}q_{1}}{\sqrt{G_{1}^{2}+G_{2}^{2}}}\text{, }\hspace{0.5cm} p_{-}=%
\frac{G_{1}p_{2}-G_{2}p_{1}}{\sqrt{G_{1}^{2}+G_{2}^{2}}}.
\end{align}
\end{subequations}%
In the representation of the two new hybrid modes, the Hamiltonian $H_{%
\text{lin}}$ can be expressed as%
\begin{eqnarray}
H_{\text{lin}}/\hbar  &=&\Delta a^{\dagger }a+\frac{\Omega _{+}}{2}(
q_{+}^{2}+p_{+}^{2}) +\frac{\Omega _{-}}{2}(q_{-}^{2}+p_{-}^{2})   \notag \\
&&+G_{q}q_{+}q_{-}+G_{p}p_{+}p_{-}+G_{+}( a^{\dagger }+a) q_{+} \notag \\
&&-G_{x}\frac{G_{1}G_{2}q_{+}^{2}-G_{1}G_{2}q_{-}^{2}}{G_{1}^{2}+G_{2}^{2}}%
\text{,}  \label{17}
\end{eqnarray}%
\begin{figure}[!t]
\center\includegraphics[width=0.48\textwidth]{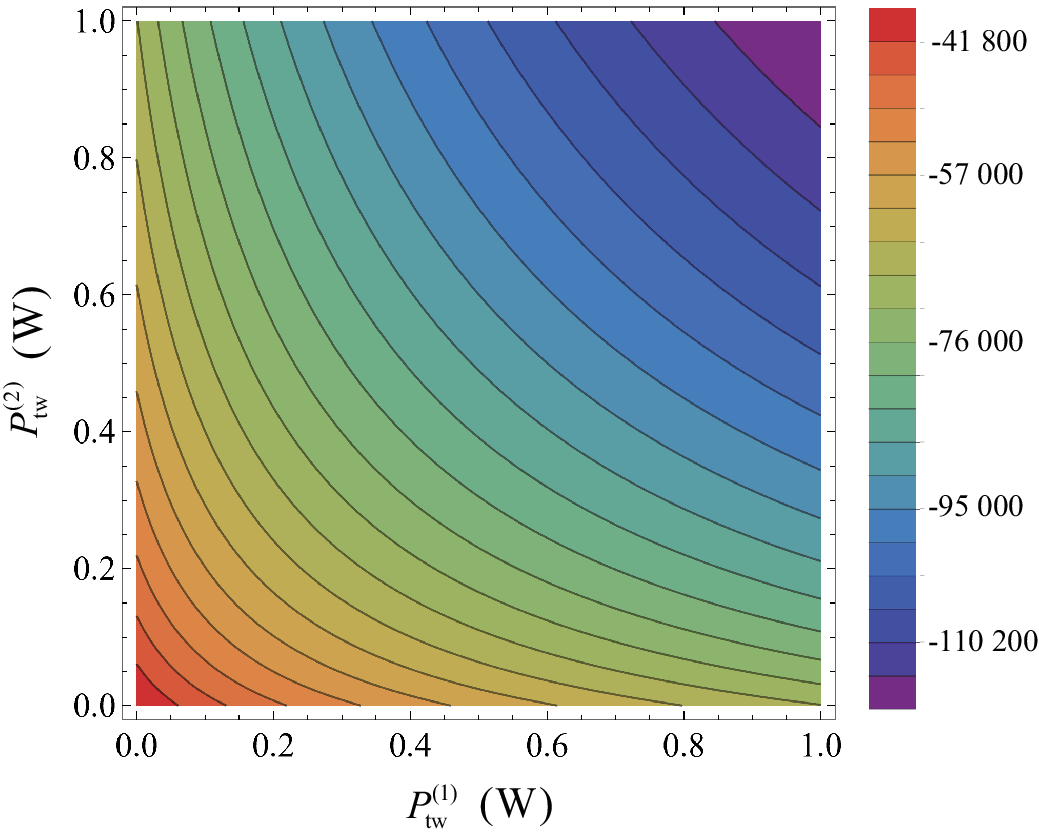}
\caption{{The particle-particle coupling strength $G_{x}$ versus
 the powers $P_{\text{tw}}^{(1)}$ and $P_{\text{tw}}^{(2)} $ of the two optical tweezers.
 Here, we take the following parameters: the radius of the two silica nanoparticles $a_{0}\approx90$ nm, the wavelength $\lambda=1064$ nm, and the numerical aperture $\text{NA}\approx0.8$.
}}
\label{modelv7}
\end{figure}%
where the resonance frequencies of the two hybrid modes are introduced by
\begin{equation}
\Omega _{+}=\frac{\Omega _{1}G_{1}^{2}+\Omega _{2}G_{2}^{2}}{%
G_{1}^{2}+G_{2}^{2}}\text{,  }\hspace{0.5cm}\Omega _{-}=\frac{\Omega _{1}G_{2}^{2}+\Omega
_{2}G_{1}^{2}}{G_{1}^{2}+G_{2}^{2}}\text{.}
\end{equation}%
In addition, the optomechanical coupling strength between the cavity mode $a$ and the
hybrid mode $q_{+}$ is given by $G_{+}=\sqrt{G_{1}^{2}+G_{2}^{2}}$, and the other two
coupling strengths between the two modes $q_{\pm}$ are given by%
\begin{subequations}
\begin{align}
G_{q} &=\frac{\left( \Omega _{2}-\Omega _{1}\right) G_{1}G_{2}-k_{x}(
G_{1}^{2}-G_{2}^{2}) }{G_{1}^{2}+G_{2}^{2}}\text{,} \\
G_{p} &=\frac{\left( \Omega _{2}-\Omega _{1}\right) G_{1}G_{2}}{%
G_{1}^{2}+G_{2}^{2}}\text{.}
\end{align}%
\end{subequations}%
It can be seen from Eq.~(\ref{17}) that, when $\Omega_{1}=\Omega_{2}$ and $G_{1}=-G_{2}$,
we have $G_{q}=G_{p}=0$, and thus the mode $q_{-}$ is decoupled from both the mode $%
q_{+}$ and the cavity mode $a$. In this case, the mode $q_{-}$ becomes a dark mode, and it cannot be cooled
via the cavity-field cooling channel. In
order to realize the simultaneous ground-state cooling of the two modes $q_{1}$ and $q_{2}$, we need to take different
optical tweezers powers, i.e., $P_{\text{tw}}^{(1)}\neq P_{\text{tw}}^{(2)}$, then the dark-mode effect is broken.

In Fig.~\ref{modelv4}(a),
we plot the final mean phonon numbers $\bar{n}_{1}$ and $\bar{n}_{2}$ for the two mechanical modes
as functions of the scaled detuning $\Delta/\Omega_{1}$ in the nondegenerate-mechanical-mode case, $\Omega _{2}\approx 0.75\Omega _{1}$.
In this case, we obtain the mechanical frequencies $\Omega_{x,y,z}/2\pi\approx(324,366,130)$ kHz for the center-of-mass motion
of the particle 1.
For this system, the nanoparticle is levitated in a vacuum, and hence it is highly isolated from the environment, resulting in a high $Q$ factor exceeding $10^{8}$~\cite{JNP2013}.
Under these parameters, the simultaneous ground-state cooling of the two mechanical modes can be realized $\left(\bar{n}_{1},\bar{n}_{2}<1\right) $. In particular, the optimal cooling of the mode $x_{j}$ for the $j$th
particle appears around the red-sideband resonances: $\Delta/\Omega_{j}\approx1$. Note that the slight shifts of the resonance point are caused by the couplings among the optical modes and two mechanical modes. We point out that the present cooling mechanism is essentially a sideband cooling. Therefore, we need to investigate the dependence of the cooling performance on the sideband-resolution condition. To this end, we plot in Fig.~\ref{modelv4}(b) the final mean phonon numbers $\bar{n}_{1}$ and $\bar{n}_{2}$ as functions of the scaled cavity-field decay rate $\kappa/\Omega_{1}$. Here we can see that the final mean phonon numbers $\bar{n}_{1}$ and $\bar{n}_{2}$ firstly decrease and then increase with the increase of the decay rate $\kappa$ (See the inset)~\cite{CPRA2008,MPRL2018}.
\begin{figure}[!t]
\center\includegraphics[width=0.48\textwidth]{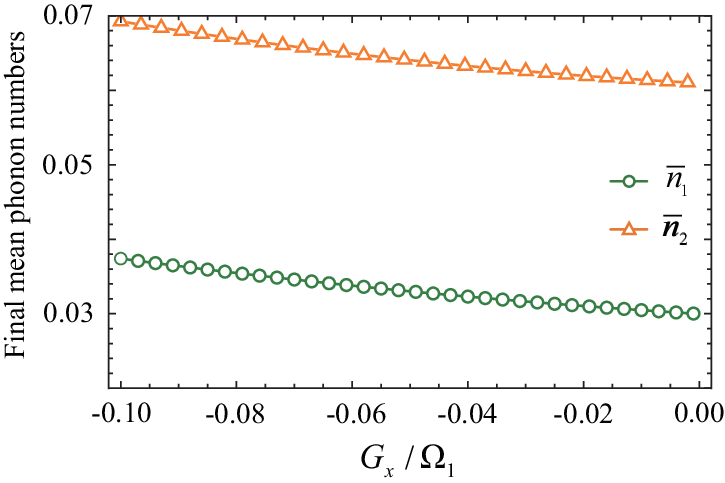}
\caption{{The final mean phonon numbers $\bar{n}_{1}$ (green solid line with
circles) and $\bar{n}_{2}$ (yellow solid line with triangles) as functions of
$G_{x}/\Omega_{1}$ when $\Delta=\Omega_{1}$ and $\kappa/\Omega_{1}=0.2$. Other parameters used are the same as those in Fig.~\ref{modelv4}
}}
\label{modelv6}
\end{figure}%

In this system, there exists a coupling between the two mechanical modes with the coupling strength $G_{x}$.
Below, we investigate how the coupling strength $G_{x}$ affects the cooling results of the two mechanical modes. Firstly, we point out that the particle-particle coupling strength used in Fig.~\ref{modelv4} is $G_{x}\approx-0.046\Omega_{1}$, where the distance between the two particles is given by $D\approx2.5\lambda$.
It can be confirmed from Fig.~\ref{modelv2} that the used parameters satisfy the far-field approximation well.
In addition, it can be seen from Eq.~(\ref{kx}) that the particle-particle coupling strength $G_{x}$ can be adjusted
by the powers $P_{\text{tw}}^{(1)}$ and $P_{\text{tw}}^{(2)}$ of the two tweezers and the distance $D$
of the two particles.
To further elucidate this point, in Fig.~\ref{modelv7} we plot the particle-particle coupling strength $G_{x}$ versus the powers $P_{\text{tw}}^{(1)}$ and $P_{\text{tw}}^{(2)}$ of the two tweezers. Figure~\ref{modelv7} exhibits that the strength $G_{x}$ increases with the increase of the two powers $P_{\text{tw}}^{(1)}$ and $P_{\text{tw}}^{(2)}$, which is consistent with the phenomenon that the optical binding force between the two levitated particles increases with the powers of the two tweezers. Moreover, we find that $G_{x}$ can be adjusted from $-41.8$~kHz to $-110.2$~kHz and the scaled coupling $G_{x}/\Omega_{1}\in[-0.1,-0.03]$.
Since the particle-particle coupling provides a channel for the exchange of thermal excitations between the two mechanical modes $x_{1}$ and $x_{2}$, it is interesting to analyze the dependence of the cooling results on the coupling strength $G_{x}$. In Fig.~\ref{modelv6}, we plot the final mean phonon numbers $\bar{n}_{1}$ and $\bar{n}_{2}$ as functions of the scaled particle-particle coupling strength $G_{x}/\Omega_{1}$. Figure~\ref{modelv6} shows that both the final mean phonon numbers $\bar{n}_{1}$ and $\bar{n}_{2}$ of
 particles 1 and 2 increase with the increase of the absolute value of the particle-particle coupling $|G_{x}|$,
 which means that the increase of the $|G_{x}|$ will
reduce the cooling efficiency of the two modes $x_{1}$ and $x_{2}$.

We mention that there exists a dilemma in the choosing of the resonance frequencies $\Omega_{j=1,2}$. As shown in Eq.~(\ref{16}), the dipole-induced coupling determines the phonon exchange between the two mechanical modes, and this effect is strong in the resonant case. To break the dark-mode effect, however, we prefer to consider two detuned mechanical modes. In our simulations, we choose detuned mechanical modes such that the dark-mode effect is broken. In this case, the dipole-induced coupling is partly suppressed but it still works.

\section{Simultaneous cooling of x- and z-direction motions}\label{Transition to four modes cooling}

In Sec.~\ref{Cooling to two modes ground-state}, we study the special case where the distance
between the two particles takes special values, resulting in the
decoupling of the cavity mode $a$ from the $z$-direction motions of the two
particles. However, when the distance between the two particles does not take these
special positions, the cavity mode $a$ will couple to the $z$-direction motions %
of the two particles. In the following, we will analyze
the cooling of both $x$- and $z$-direction motions of the two particles in this case.

For a general case, the system is described by the Hamiltonian~(\ref{Htot}).
Here, there exist nonlinear couplings between the cavity mode and the $x$-direction
motions of the two particles. In particular, we consider the case where the driving
[the $\tilde{\Omega}$ term in Eq.~(\ref{Htot})] of the cavity mode is strong enough, then we
can linearize the system and obtain the linearized Langevin equations as
\begin{eqnarray}
\mathbf{\dot{u}}^{\prime }\left( t\right)
=\mathbf{A^{\prime }}\mathbf{u^{\prime }}\left( t\right) +\mathbf{N}^{\prime }\left( t\right)\text{.} \label{29}
\end{eqnarray}%
Here the fluctuation operator vector is defined by
\begin{eqnarray}
\mathbf{u}^{\prime }\left( t\right) &=&(
\delta q_{1x} , \delta p_{1x} , \delta q_{2x} , \delta p_{2x} ,\delta q_{1z}
, \delta p_{1z}, \delta q_{2z} , \delta p_{2z} ,\notag \\
&& \delta X , \delta Y%
) ^{\text{T}} \text{,}
\end{eqnarray}%
where the mechanical quadratures are introduced as $\sqrt{2}\delta q_{jx}=x_{j}/x_{j\text{,zpf}}$, $\sqrt{2}\delta p_{jx}=P_{jx}/p_{jx,\text{zpf}}$, $\sqrt{2}\delta q_{jz}=z_{j}/z_{j\text{,zpf}}$, and $\sqrt{2}\delta p_{jz}=P_{jz}/p_{jz\text{,zpf}}$ with $j=1,2$. In Eq.~(\ref{29}), the noise operator vector is defined by
\begin{eqnarray}
\mathbf{N}^{\prime }\left(t\right)&=&(0,f_{\text{th}}^{(1x)}(t),
0,f_{\text{th}}^{(2x)}(t),0,f_{\text{th}}^{(1z)}(t),0,f_{\text{th}}^{(2z)}(t), \notag \\
&& \sqrt{%
2\kappa }\delta X_{\text{in}}\left( t\right) ,\sqrt{2\kappa }\delta Y_{%
\text{in}}\left( t\right))^{\text{T}},
\end{eqnarray}%
and the coefficient matrix is given by
\begin{widetext}
\begin{equation}
\mathbf{A^{\prime }=}\left(
\begin{array}{cccccccccc}
0 & \tilde{\omega}_{1x} & 0 & 0 & 0 & 0 & 0 & 0 & 0 & 0 \\
-\tilde{\omega}_{1x} & -\gamma _{1x} & G_{x} & 0 & 0 & 0 & 0 & 0 & -\sqrt{2%
}A_{1}  & \sqrt{2}B_{1}  \\
0 & 0 & 0 & \tilde{\omega}_{2x} & 0 & 0 & 0 & 0 & 0 & 0 \\
G_{x} & 0 & -\tilde{\omega}_{2x} & -\gamma _{2x} & 0 & 0 & 0 & 0 & -\sqrt{2%
}A_{2}  & \sqrt{2}B_{2}  \\
0 & 0 & 0 & 0 & 0 & \tilde{\omega}_{1z} & 0 & 0 & 0 & 0 \\
0 & 0 & 0 & 0 & -\tilde{\omega}_{1z} & -\gamma _{1z} & G_{z} & 0 & -\sqrt{2%
}C_{1} & -\sqrt{2}D_{1} \\
0 & 0 & 0 & 0 & 0 & 0 & 0 & \tilde{\omega}_{2z} & 0 & 0 \\
0 & 0 & 0 & 0 & G_{z} & 0 & -\tilde{\omega}_{2z} & -\gamma _{2z} & -\sqrt{2%
}C_{2} & -\sqrt{2}D_{2} \\
-\sqrt{2}B_{1}  & 0 & -\sqrt{2}B_{2}
  & 0 & -\sqrt{2}D_{1} & 0 & -\sqrt{2}D_{2} & 0 & -\kappa
& +\tilde{\Delta} \\
-\sqrt{2}A_{1}  & 0 & -\sqrt{2}A_{2}
& 0 & -\sqrt{2}C_{1} & 0 & -\sqrt{2}C_{2} & 0 & -\tilde{\Delta} & -\kappa
\end{array}%
\right). \label{A'}
\end{equation}
\end{widetext}
In Eq.~(\ref{A'}), we have defined $G_{x}=2k_{x}x_{1\text{,zpf}}x_{2\text{,zpf}}/\hbar ,$ $G_{z}=2k_{z}z_{1%
\text{,zpf}}z_{2\text{,zpf}}/\hbar ,$ $G_{x_{j}}=\sqrt{2}\tilde{g}_{x_{j}}x_{j\text{,zpf}}$,
$G_{z_{j}}=i\sqrt{2}\tilde{g}_{z_{j}}z_{j\text{,zpf}}$, and $G_{ax_{j}}=\sqrt{2}\tilde{g}_{ax_{j}}\langle
a^{\dagger }\rangle x_{j\text{,zpf}}$, then the linearized optomechanical-coupling strengths are given by
$\tilde{G}_{jx}=G_{x_{j}}+G_{ax_{j}}$ and $\tilde{G}_{jz}=G_{z_{j}}$. These complex coupling
strengths can be divided into real and imaginary parts, namely, $%
\tilde{G}_{jx}=A_{j}+iB_{j}$ and $\tilde{G}_{jz}=C_{j}+iD_{j}$, where $A_{j}$, $B_{j}$, $C_{j}$, and $D_{j}$ have been introduced in Eq.~(\ref{A'}).
\begin{figure}[!t]
\center\includegraphics[width=0.48\textwidth]{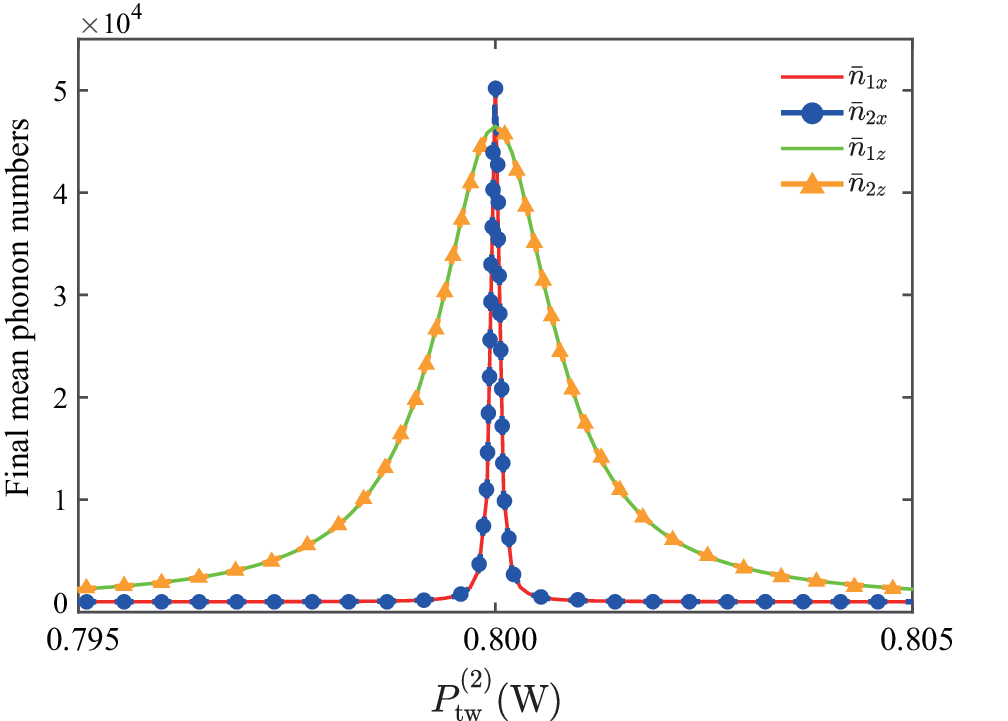}
\caption{{The final mean phonon numbers $\bar{n}_{1x}$ (red line), $\bar{n}_{2x}$
(blue dashed line with dots), $\bar{n}_{1z}$ (green line), and $\bar{n}_{2z}$ (yellow dashed line with triangles) in
the four mechanical modes versus the power $P_{\text{tw}}^{(2)}
$ of the tweezer 2. Other parameters used are $P_{\text{tw}}^{(1)}=0.8$ W, $\bar{n}_{1x\text{,th}}=\bar{n}_{2x\text{,th}}=\bar{n}_{1z\text{,th}}=\bar{n}_{2z\text{,th}}=10^{5}$,
$\gamma _{1x}/\tilde{\omega}_{1x}=\gamma _{1z}/\tilde{\omega}_{1x}= \gamma_{2x}/\tilde{\omega}_{1x}=\gamma _{2z}/\tilde{\omega}_{1x}= 0.5 \times 10^{-8}$, and $\kappa/\tilde{\omega}_{1x}=0.2$.}}
\label{modelv8}
\end{figure}%

These linearized coupling strengths depend on the semi-classical motion, which are governed by the semi-classical equations of motion.
In the steady-state case, the average values of the
system operators can be obtained as%
\begin{subequations}
\begin{align}
\left\langle a\right\rangle =&\frac{i\tilde{\Omega} ^{\ast
}+iG_{x_{1}}^{\ast }\left\langle x_{1}\right\rangle+iG_{x_{2}}^{\ast }\left\langle x_{2}\right\rangle+iG_{z_{1}}^{\ast
}\left\langle z_{1}\right\rangle+iG_{z_{2}}^{\ast }\left\langle z_{2}\right\rangle}{-i\tilde{\Delta}-\kappa },
\\
\left\langle x_{1}\right\rangle =&\frac{-G_{ax_{1}}%
\left\langle a\right\rangle  +G_{x}\left\langle x_{2}\right\rangle-\tilde{R}_{1}-2\text{Im}%
\left[ G_{x_{1}}\left\langle a\right\rangle \right] }{\tilde{\omega}_{1x}}, \\
\left\langle x_{2}\right\rangle =&\frac{-G_{ax_{2}}%
\left\langle a\right\rangle  +G_{x}\left\langle x_{1}\right\rangle +\tilde{R}_{2}-2\text{Im}%
\left[ G_{x_{2}}\left\langle a\right\rangle \right] }{\tilde{\omega}_{2x}}, \\
\left\langle z_{1}\right\rangle =&\frac{G_{z}\left\langle z_{2}\right\rangle-2\text{%
Im}\left[ G_{z_{1}}\left\langle a\right\rangle \right] }{\tilde{\omega}_{1z}}, \\
\left\langle z_{2}\right\rangle =&\frac{G_{z}\left\langle z_{1}\right\rangle-2\text{%
Im}\left[ G_{z_{2}}\left\langle a\right\rangle \right] }{\tilde{\omega}_{2x}},
\end{align}%
\end{subequations}%
where $\tilde{\Delta}=\Delta ^{\prime }+\tilde{g}_{ax_{1}}x_{1\text{,zpf}%
}\langle x_{1}\rangle +\tilde{g}_{ax_{2}}x_{2\text{,zpf}}\langle x_{2}\rangle$ and $\tilde{R}_{j}=\sqrt{2}\tilde{R}x_{j\text{,zpf}} $. %
Based on the linearized Langevin equations, we can derive an effective Hamiltonian to describe the linearized dynamics of the system as
\begin{eqnarray}
H_{\text{lin}}^{\prime }/\hbar  &=&\tilde{\Delta}\delta a^{\dagger }\delta a+\sum_{l}[\frac{\tilde{%
\omega}_{l}}{2}( \delta q_{l}^{2}+\delta p_{l}^{2}) +( \tilde{G}_{l}\delta a+%
\tilde{G}_{l}^{\ast }\delta a^{\dagger }) \delta q_{l}]  \notag \\
&&-G_{x}\delta q_{1x}\delta q_{2x}-G_{z}\delta q_{1z}\delta q_{2z}\text{,} \label{GHlin}
\end{eqnarray}%
where $l= 1x\text{, }2x\text{, }1z\text{, }2z$.

It can be seen from Eq.~(\ref{GHlin}) that, the cavity mode $\delta a$ is coupled to the four modes $\delta q_{1x}$, $\delta q_{1z}$, $\delta q_{2x}$, and $\delta q_{1z}$. Meanwhile, the modes $\delta q_{1x}$ and $\delta q_{1z}$ are coupled to the modes $\delta q_{2x}$ and $\delta q_{2z}$, respectively. To investigate the dependence of the cooling performance of the four mechanical modes on the powers of the two tweezers, in Fig.~\ref{modelv8}, we plot the final mean phonon number $\bar{n}_{1x}$, $\bar{n}_{1z}$, $\bar{n}_{2x}$, and $\bar{n}_{2z}$ as functions of $P_{\text{tw}}^{(2)}$  when $P_{\text{tw}}^{(1)}=0.8$~W.
Here, we can see that all the modes cannot be cooled around the identical power point $P_{\text{tw}}^{(1)}\approx P_{\text{tw}}^{(2)}$. The phenomenon can be explained based on the dark-mode effect. In this five-mode system, when $P_{\text{tw}}^{(1)}=P_{\text{tw}}^{(2)}$, we have $\tilde{\omega}_{1x}=\tilde{\omega}_{2x}$ and $\tilde{\omega}_{1z}=\tilde{\omega}_{2z}$. Meanwhile, the coupling strengths satisfy the relations: $G_{x_{1}}=-G_{x_{2}}$, $G_{ax_{1}}=-G_{ax_{2}}$, and $G_{z_{1}}=G_{z_{2}}$. To analyze the dark-mode effect, we introduce the creation and annihilation operators $b_{l}^{\dagger}=(\delta q_{l}-i\delta p_{l})/\sqrt{2}$ and $b_{l}=(\delta q_{l}+i\delta p_{l})/\sqrt{2}$ of these mechanical modes. We further define four hybrid mechanical modes as
\begin{subequations}
\begin{align}
B_{1+}=&\frac{\tilde{G}_{1x}b_{1x}+\tilde{G}_{2x}b_{2x}}{\sqrt{\left|\tilde{G}_{1x}\right|^{2}+\left|\tilde{G}_{2x}\right|^{2}}}\text{, }\hspace{0.5cm}%
B_{1-}=\frac{\tilde{G}_{2x}^{*}b_{1x}-\tilde{G}_{1x}^{*}b_{2x}}{\sqrt{\left|\tilde{G}_{1x}\right|^{2}+\left|\tilde{G}_{2x}\right|^{2}}}\text{, } \\
B_{2+}=&\frac{\tilde{G}_{1z}b_{1z}+\tilde{G}_{2z}b_{2z}}{\sqrt{\left|\tilde{G}_{1z}\right|^{2}+\left|\tilde{G}_{2z}\right|^{2}}}\text{, }\hspace{0.5cm}%
B_{2-}=\frac{\tilde{G}_{2z}^{*}b_{1z}-\tilde{G}_{1z}^{*}b_{2z}}{\sqrt{\left|\tilde{G}_{1z}\right|^{2}+\left|\tilde{G}_{2z}\right|^{2}}}\text{. }
\end{align}%
\end{subequations}%
\begin{figure}[t]
\center\includegraphics[width=0.48\textwidth]{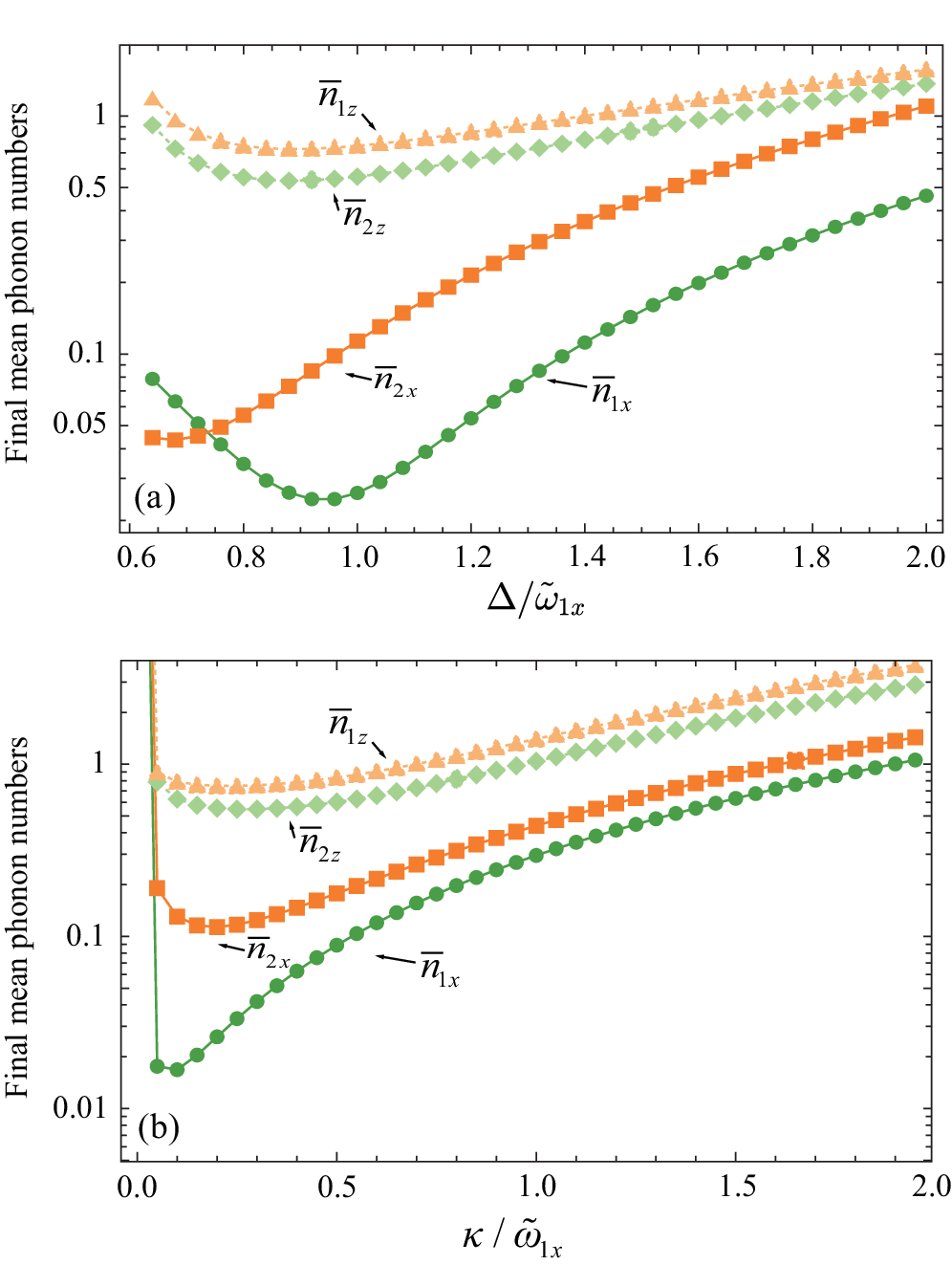}
\caption{{The final mean phonon numbers $\bar{n}_{1x}$ (green line with dots), $\bar{n}_{2x}$
(orange line with squares), $\bar{n}_{1z}$ (light green dashed line with rhombuses), and $\bar{n}_{2z}$ (yellow dashed line with triangles) versus the effective driving detuning $\Delta/\tilde{\omega}_{1x}
$. Other parameters used are $\bar{n}_{1x\text{,th}}=\bar{n}_{2x\text{,th}}=\bar{n}_{1z\text{,th}}=\bar{n}_{2z\text{,th}}=10^{5}$,  $\tilde{\omega}_{2x}/\tilde{\omega}_{1x}\approx0.75$
, $\tilde{\omega}_{1z}/\tilde{\omega}_{1x}\approx0.41$, $\tilde{\omega}_{2z}/\tilde{\omega}_{1x}\approx0.31$, $G_{x}/\tilde{\omega}_{1x}\approx-0.02$,
$G_{z}/\tilde{\omega}_{1x}=-0.03$, $G_{x_{1}}/\tilde{\omega}_{1x}\approx-0.1$, $G_{x_{2}}/\tilde{\omega}_{1x}\approx-0.09$, $\gamma _{1x}/\tilde{\omega}_{1x}=\gamma _{1z}/\tilde{\omega}_{1x}= \gamma
_{2x}/\tilde{\omega}_{1x}=\gamma _{2z}/\tilde{\omega}_{1x}= 0.5 \times 10^{-8}$, and $\kappa/\tilde{\omega}_{1x}=0.2$.}}
\label{modelv5}
\end{figure}%
In the hybrid-mode representation, the Hamiltonian~(\ref{GHlin}) can be re-expressed as a new form, which is not presented here because of its complicated form. By analyzing the Hamiltonian in the hybrid-mode representation, we find that the dark-mode effect appears under the conditions $\tilde{\omega}_{1x(z)}=\tilde{\omega}_{2x(z)}$ and $\tilde{G}_{1x(z)}^{2}=\tilde{G}_{2x(z)}^{2}$. Accordingly, in the identical power case under consideration, the $x(z)$ modes of the two particles have
the same frequency $\tilde{\omega}_{1x(z)}=\tilde{\omega}_{2x(z)}$
and absolute value of coupling strength $|\tilde{G}_{1x(z)}|=|\tilde{G}_{2x(z)}|$,
 and then the Hamiltonian is reduced to
\begin{eqnarray}
H_{\text{lin}}^{\prime }/\hbar &=&\Delta \delta a^{\dagger }\delta a+\sum_{j=1,2}\tilde{\omega}%
_{j}(B_{j+}^{\dagger }B_{j+}+B_{j-}^{\dagger }B_{j-}) \notag \\
&&+\sum_{j=1,2}(-1)^{j}[(\xi_{j} B_{j+}^{\dagger}+\xi_{j}^{*} B_{j+})^{2} \notag -(\xi_{j}^{*}B_{j-}^{\dagger}+\xi_{j} B_{j-})^{2}] \notag \\
&&+[\tilde{G}_{1x}(\zeta_{1} B_{1+}+\zeta_{1}^{*} B_{1+}^{\dagger} )a+\tilde{G}_{1z}(\zeta_{2} B_{2+}+\zeta_{2}^{*} B_{2+}^{\dagger} )a \notag \\
&&+\text{H.c.}],
\label{30}
\end{eqnarray}
where we have ignored the constant term. In Eq.~(\ref{30}), the normalized resonance frequencies are
$\tilde{\omega}_{1}=\tilde{\omega}_{1x}$ and $\tilde{\omega}_{2}=\tilde{\omega}_{1z}$,
and other parameters used are defined as $\zeta_{1}=G_{x}/2\tilde{G}_{1x}^{*}$, $\zeta_{2}=G_{z}/2\tilde{G}_{1z}$,
$\xi_{1}=\sqrt{2}\tilde{G}_{1x}^{*}/|\tilde{G}_{1x}|$, and $\xi_{2}=\sqrt{2}\tilde{G}_{1z}^{*}/|\tilde{G}_{1z}|$. We can see from Eq.~(\ref{30}) that the two modes $B_{1-}$ and $B_{2-}$ are decoupled from other modes, and hence become the dark modes. Therefore, the cooling of the four mechanical modes will be significantly suppressed.

In order to break the dark-mode effect, we need to consider the
case $P_{\text{tw}}^{(1)}\neq P_{\text{tw}}^{(2)}$, then we have $|g_{z}|>|g_{x}|$ and $\tilde{\omega}_{z}<\tilde{\omega}_{x}$ for our parameters. In the dark-mode-breaking case, we want to investigate the influence of the detuning on the cooling performance for the motions. In Fig.~\ref{modelv5}(a), we plot the final mean phonon numbers $\bar{n}_{1x}$, $\bar{n}_{2x}$, $\bar{n}_{1z}$, and $\bar{n}_{2z}$ in the four mechanical modes versus the scaled detuning $\Delta/\tilde{\omega}_{1x}$.
 We find in Fig.~\ref{modelv5}(a) that the simultaneous ground-state cooling of the four mechanical
modes can be realized. In particular, the cooling performance of the $x$-direction motions are better than those of the two modes $z_{1}$ and $z_{2}$. Moreover, we can see from Fig.~\ref{modelv5}(a) that, for the two modes $x_{1}$ and $x_{2}$, the optimal cooling appears respectively around $\Delta/\tilde{\omega}_{1x}\approx1$ and $\Delta/\tilde{\omega}_{2x}\approx0.75$, corresponding to the red-sideband resonances. In this case, the effective mode temperatures of the $x$-direction motions for the two nanoparticles are cooled to $T_{x}\sim5\times10^{-3}$ mK (corresponding to $\tilde{\omega}_{x}\sim2.4$ MHz), and the two $z$-mode mechanical oscillations are simultaneously cooled to $T_{z}\sim7\times10^{-3}$ mK (corresponding to $\tilde{\omega}_{z}\sim0.9$ MHz). We also investigate the dependence of the final mean phonon numbers $\bar{n}_{1x}$, $\bar{n}_{2x}$, $\bar{n}_{1z}$, and $\bar{n}_{2z}$ on the scaled decay rate $\kappa/\tilde{\omega}_{1x}$ for the cavity field, as shown by Fig.~\ref{modelv5}(b). Here, we can see that the simultaneous ground-state cooling of the four modes can be realized. In particular, the cooling performances of the two $x$-direction modes $x_{1}$ and $x_{2}$ are better than those of the two $z$-direction modes $z_{1}$ and $z_{2}$. Meanwhile, we find that the main cooling region exists in the resolved-sideband regime.

\section{Discussions and Conclusion}\label{conclusion}
Finally, we present some discussions concerning the decoherence effect in this scheme. In reality, there exist some noise processes such as recoil heating, photon absorption, and laser phase noise. These noises will inevitably cause some system decoherence. In this work, we did not conduct quantitative analyses concerning the effect of these noises separately. We only consider that the cavity mode and each mechanical mode are coupled to a vacuum bath and a heat bath, respectively. We also consider a real dielectric constant (i.e., no absorption effects are taken into account). In addition, the heating noise originates from two contributions: one is the heating effect caused by the collision of the nanoparticles with surrounding gas molecules, the other one is the recoil heating caused by the collision of the nanoparticles with the photons escaped from the optical trap. According to Ref.~\cite{VPRL2016}, the ratio between the decoherence rate induced by the photon shot noise and thermal decoherence induced by the gas collision increases as the decrease of the vacuum pressure. In this work, we consider the case in which the decoherence induced by
gas collisions dominates the heating processes, and hence the decoherence of the system can be described by the Langevin equations.

We should mention that the cooling performance depends on the initial thermal phonon numbers in these modes. In our simulations, we consider the initial phonon number $\bar{n}_{\text{th}}=10^{5}$. We also conduct the simulations for a larger initial $\bar{n}_{\text{th}}$, for example $\bar{n}_{\text{th}}=10^{7}$ corresponding to a room temperature for typical mechanical frequency. We find that the ground-state cooling for the $x$ modes of the two nanoparticles can be realized at $\bar{n}_{\text{th}}=10^{7}$. However, the ground-state cooling of the $x$ and $z$ modes of the two nanoparticles cannot be realized, but the final mean phonon numbers can be reduced to be smaller than $10$ with the cooperativity $C_{Q}\approx4g^{2}/(\kappa \gamma \bar{n}_{\text{th}})\approx4$.

We note that the normal-mode splitting caused by a strong optical binding~\cite{JS2022} can limit the cooling efficiency. In particular, when the two coupled mechanical modes are working in the ultrastrong-coupling regime (namely the coupling strength is considerably comparable to the mechanical resonance frequency), the two mechanical oscillators should be treated as a whole system. Namely, the mechanical modes are no longer individually coupled to the thermal bath. Instead, each of the two hybrid modes will be coupled to two environments, and hence the equations of motion governing the evolution of the two coupled mechanical modes should be re-derived in the dressed-mode representation. In other word, we need to consider this model in the normal-mode representation when the two coupled mechanical modes work in the ultrastrong-coupling regime. With the coupling between the two mechanical modes in the weak- and strong-coupling regimes, we can safely work in the bare-mode representation of the two mechanical modes. In our simulations, we consider this coupling strength $G_x\approx-0.045\Omega_{1}$ in the strong-coupling regime. Therefore, our treatment and calculations are valid for our used parameters.

In conclusion, we have developed a theoretical model for
describing the simultaneous ground-state cooling of the motions of
two levitated nanoparticles trapped in a cavity via coherent scattering.
We have found that, different from the single-levitated particle case,
the scattered light will induce the mechanical effect between the
particles, which shifts the equilibrium position
of the particles and causes the coupling between two particles. We have derived the
Hamiltonian of the system and analyze the interactions in various cases. When the two nanoparticles are located at the nodes of the
cavity, the system is reduced to a three-mode loop-coupled model, in which the cavity mode is coupled to the $x$-direction motional modes of the two particles, and the two $x$ modes are coupled with each other via the position-position coupling.
In this case, we have found that the dark-mode effect appears when two tweezers have
the same power, and then the effective cooling of the two mechanical oscillations is suppressed.
In particular, the simultaneous ground-state cooling of the $x$-direction motion of the
two particles can be realized by breaking the dark-mode effect. In addition, when the particles are not
placed at the nodes, the system is reduced to a five-mode model, in which both the $x$- and $z$-direction motions are
coupled to the cavity mode, and there exist both the $x$-$x$ coupling and $z$-$z$ coupling between the two mechanical modes. In this case, we have also found that the dark-mode effect exists in the identical-power case, and that both the $x$- and $z$-direction motions can be significantly cooled by breaking the dark-mode effect.
This work paves the way to quantum manipulation of multiple levitated nanoparticles.

\begin{acknowledgments}
J.-Q.L. was supported in part by National Natural Science Foundation of China (Grants No.~12175061, No.~12247105, and No.~11935006), the Science and Technology Innovation Program of Hunan Province (Grant No.~2021RC4029), and Hunan Provincial Major Science and Technology Program (Grant No.~2023ZJ1010).
\end{acknowledgments}

\appendix*
\begin{widetext}
\section{Expansions of these interaction Hamiltonians described by Eqs.~(\ref{crossterm})} \label{appendix}
In this Appendix, we present the derivation of these interaction Hamiltonians given by Eqs.~(\ref{crossterm}).
We start by substituting Eqs.~(\ref{Ej0}) and~(\ref{15}) into  Eq.~(\ref{12}), and obtain
\begin{eqnarray}
\hat{H}_{\text{tw-Gtw}}^{\left( 1\right) } &=&-\alpha \bm{\mathcal{E}}_{\text{tw1}%
}(\boldsymbol{\hat{R}}_{1},t)\cdot\text{Re}[e^{ik_{\text{tw}}\hat{R}_{0}}\eta
_{f_{\text{tw}}}(D/\hat{R}_{0})\overleftrightarrow{\mathbf{M}}_{f%
}(\boldsymbol{\hat{R}}_{0})\cdot\boldsymbol{E}_{\text{tw2}}(\boldsymbol{\hat{R}}_{2},t)]
\notag \\
&=&-\alpha \left[ \frac{1}{2}E_{10}(\boldsymbol{\hat{R}}_{1})e^{-ik_{\text{%
tw}}\hat{Z}_{1}}e^{-i\omega _{\text{tw}}t}+\frac{1}{2}E_{10}(\boldsymbol{%
\hat{R}}_{1})e^{ik_{\text{tw}}\hat{Z}_{1}}e^{i\omega _{\text{tw}}t}\right]%
\boldsymbol{e}_{\text{tw}}^{(1)}  \cdot \text{Re}[ e^{ik_{\text{tw}}\hat{R}_{0}}\eta _{f_{\text{tw}}}(D/%
\hat{R}_{0})\overleftrightarrow{\mathbf{M}}_{f}(\boldsymbol{\hat{%
R}}_{0})\cdot \boldsymbol{E}_{20}(\boldsymbol{\hat{R}}_{2})e^{-ik_{\text{tw}}%
\hat{Z}_{2}}e^{-i\omega _{\text{tw}}t}\boldsymbol{e}_{\text{tw}}^{(2)}]
\notag \\
&\approx &-\frac{1}{2}\alpha \eta _{f_{\text{tw}}}(D/\hat{R}_{0})[\frac{1}{2}%
E_{10}(\boldsymbol{\hat{R}}_{1})E_{20}(\boldsymbol{\hat{R}}_{2})e^{-ik_{%
\text{tw}}\hat{R}_{0}}e^{-ik_{\text{tw}}( \hat{Z}_{1}-\hat{Z}%
_{2}) }  +\frac{1}{2}E_{10}(\boldsymbol{\hat{R}}_{1})E_{20}(\boldsymbol{\hat{R}}%
_{2})e^{ik_{\text{tw}}\hat{R}_{0}}e^{ik_{\text{tw}}( \hat{Z}_{1}-\hat{Z}%
_{2}) }]\boldsymbol{e}_{\text{tw}}^{(1)}\cdot \overleftrightarrow{%
\mathbf{M}}_{f}(\boldsymbol{\hat{R}}_{0})\cdot \boldsymbol{e}_{%
\text{tw}}^{(2)}  \notag \\
&=&-\frac{1}{2}\alpha \eta _{f_{\text{tw}}}(D/\hat{R}_{0})E_{10}(\boldsymbol{%
\hat{R}}_{1})E_{20}(\boldsymbol{\hat{R}}_{2})\cos (-k_{\text{tw}}\hat{Z}%
_{0}-k_{\text{tw}}\hat{R}_{0})\boldsymbol{e}_{\text{tw}}^{(1)}\cdot
\overleftrightarrow{\mathbf{M}}_{f}(\boldsymbol{\hat{R}}%
_{0})\cdot \boldsymbol{e}_{\text{tw}}^{(2)}, \label{A1}
\end{eqnarray}%
where we neglect the fast-oscillating terms by performing the rotating-wave approximation. Similarly, we can obtain the Hamiltonian%
\begin{eqnarray}
\hat{H}_{\text{tw-Gtw}}^{\left( 2\right) } &=&-\alpha \bm{\mathcal{E}}_{\text{tw2}%
}(\boldsymbol{\hat{R}}_{2},t)\cdot\text{Re}[e^{ik_{\text{tw}}\hat{R}_{0}}\eta
_{f_{\text{tw}}}(D/\hat{R}_{0})\overleftrightarrow{\mathbf{M}}_{f%
}(\boldsymbol{\hat{R}}_{0})\cdot \boldsymbol{E}_{\text{tw1}}(\boldsymbol{\hat{R}}_{0},t)]
\notag \\
&\approx &-\frac{1}{2}\alpha \eta _{f_{\text{tw}}}(D/\hat{R}_{0})E_{20}(%
\boldsymbol{\hat{R}}_{2})E_{10}(\boldsymbol{\hat{R}}_{1})\cos (k_{\text{tw}}%
\hat{Z}_{0}-k_{\text{tw}}\hat{R}_{0})\boldsymbol{e}_{\text{tw}}^{(2)}\cdot \overleftrightarrow{%
\mathbf{M}}_{f}(\boldsymbol{\hat{R}}_{0})\cdot \boldsymbol{e}_{%
\text{tw}}^{(1)}. \label{A2}
\end{eqnarray}%
\newline
Based on Eqs.~(\ref{A1}) and~(\ref{A2}), the Hamiltonian $\hat{H}_{\text{tw-Gtw}}$ given in Eq.~(\ref{HTGT1}) can be derived. Since the $j$th particle is trapped near the focus of the $j$th tweezer, we can approximate the electric field by its expansion near the foci $r_{10} = (x_{10},0,0)$ and $r_{20}=(x_{20},0,0)$ of the two tweezers. The approximate Hamiltonian expanded up to the second order of the center-of-mass displacements is given by Eq.~(\ref{18}).

In the rotating frame with respect to $U=\exp(-i\omega_{\text{tw}}\hat{a}^{\dagger}\hat{a}t)$, the electric field operator of the cavity field can be expressed as $\boldsymbol{\hat{E}}_{\text{cav}}(\boldsymbol{\hat{R}}_{j})=\epsilon _{\text{cav}}\cos (k\hat{X}_{j})(\hat{a}^{\dagger}e^{i\omega _{\text{tw}}t}+\hat{a}e^{-i\omega _{\text{tw}}t})\boldsymbol{e}_{\text{cav}}$. By substituting $\boldsymbol{\hat{E}}_{\text{cav}}(\boldsymbol{\hat{R}}_{j})$ into Eq.~(\ref{13}) for $j=1$, we obtain
\begin{eqnarray}
\hat{H}_{\text{cav-Gcav}}^{\left( 1\right) } &=&-\alpha \boldsymbol{\hat{E}}_{%
\text{cav}}(\boldsymbol{\hat{R}}_{1})\cdot \text{Re}[e^{ik\hat{R}_{0}}\eta
_{f}(D/\hat{R}_{0})\overleftrightarrow{\mathbf{M}}_{f}(\boldsymbol{\hat{R}}%
_{0})\cdot \boldsymbol{\hat{E}}_{\text{cav}}(\boldsymbol{\hat{R}}_{2})]
\notag \\
&=&-\alpha \epsilon _{\text{cav}}\cos (k\hat{X}_{1})(\hat{a}^{\dagger
}e^{i\omega _{\text{tw}}t}+\hat{a}e^{-i\omega _{\text{tw}}t})\boldsymbol{e}_{%
\text{cav}}\cdot \text{Re}[e^{ik\hat{R}_{0}}\eta _{f}(D/\hat{R}_{0})%
\overleftrightarrow{\mathbf{M}}_{f}(\boldsymbol{\hat{R}}_{0})\cdot \epsilon
_{\text{cav}}\cos (k\hat{X}_{2})(\hat{a}^{\dagger }e^{i\omega _{\text{tw}}t}+%
\hat{a}e^{-i\omega _{\text{tw}}t})\boldsymbol{e}_{\text{cav}}]  \notag \\
&\approx &-2\alpha \epsilon _{\text{cav}}^{2}\eta _{f}(D/\hat{R}_{0})\cos (k%
\hat{X}_{1})\cos (k\hat{X}_{2})\cos (k\hat{R}_{0})(\hat{a}^{\dagger }\hat{a}+%
1/2)\boldsymbol{e}_{\text{cav}}\cdot \overleftrightarrow{\mathbf{M}}%
_{f}(\boldsymbol{\hat{R}}_{0})\cdot \boldsymbol{e}_{\text{cav}}. \label{A3}
\end{eqnarray}%
Similarly, we can obtain%
\begin{eqnarray}
\hat{H}_{\text{cav-Gcav}}^{\left( 2\right) }&=&-\alpha \boldsymbol{\hat{E}}_{\text{cav}}(\boldsymbol{\hat{R}}_{2})\cdot
\text{Re}[e^{ik\hat{R}_{0}}\eta _{f}(D/\hat{R}_{0})\overleftrightarrow{%
\mathbf{M}}_{f}(\boldsymbol{\hat{R}}_{0})\cdot \boldsymbol{\hat{E}}_{\text{%
cav}}(\boldsymbol{\hat{R}}_{1})]  \notag \\
&\approx &-2\alpha \epsilon _{\text{cav}}^{2}\eta _{f}(D/\hat{R}_{0})\cos (k%
\hat{X}_{2})\cos (k\hat{X}_{1})\cos (k\hat{R}_{0})(\hat{a}^{\dagger }\hat{a}+%
1/2)\boldsymbol{e}_{\text{cav}}\cdot \overleftrightarrow{\mathbf{M}}%
_{f}(\boldsymbol{\hat{R}}_{0})\cdot \boldsymbol{e}_{\text{cav}}\text{.} \label{A4}
\end{eqnarray}%
In terms of Eqs.~(\ref{A3}) and~(\ref{A4}), we obtain the expression of $\hat{H}_{\text{cav-Gcav}}$, which has been presented in Eq.~(\ref{HCGC1}). By expanding the Hamiltonian with respect to the center-of-mass displacements, we reach the approximate Hamiltonian
\begin{eqnarray}
\hat{H}_{\text{cav-Gcav}} &\approx&-4\alpha \epsilon _{\text{cav}}^{2}\eta _{f}\cos (kx_{10})\cos
(kx_{20})\cos \left( kD\right) \hat{a}^{\dagger }\hat{a} \notag\\
&&+4\alpha \epsilon _{\text{cav}}^{2}\eta _{f}\left\{
\lbrack k\sin \left( kD\right) +\cos ( kD)/D ]\cos
\left( kx_{10}\right) \cos \left( kx_{20}\right)
+k\sin \left( kx_{10}\right) \cos \left( kx_{20}\right) \cos \left(
kD\right)
\right\}  \hat{x}_{1}(\hat{a}^{\dagger }\hat{a}+1/2) \notag\\
&&-4\alpha \epsilon _{\text{cav}}^{2}\eta _{f}\left\{
\lbrack k\sin \left( kD\right) +\cos ( kD)/D ]\cos
\left( kx_{10}\right) \cos \left( kx_{20}\right)
-k\cos \left( kx_{10}\right) \sin \left( kx_{20}\right) \cos \left(
kD\right)
\right\}\hat{x}_{2}(\hat{a}^{\dagger }\hat{a}+1/2) \notag\\
&&+\mathcal{O}(\hat{Q}^{2}_{j}), \label{A5}
\end{eqnarray}%
where $\hat{Q}_{j}=\hat{x}_{j},\hat{y}_{j},\hat{z}_{j}$. By keeping the terms up to the first order of the center-of-mass displacements and taking $x_{10}=D/2$ and $x_{20}=-D/2$, Eq.~(\ref{A5}) is reduced to Eq.~(\ref{HCGC}).

Substituting the tweezer fields and the cavity field into Eq.~(\ref{Htgc}), we can obtain the Hamiltonian
\begin{eqnarray}
\hat{H}_{\text{tw-Gcav}}^{\left( 1\right) } &=&-\alpha \bm{\mathcal{E}}_{\text{tw1}%
}(\boldsymbol{\hat{R}}_{1},t)\cdot \text{Re}[e^{ik\hat{R}_{0}}\eta _{f}(D/%
\hat{R}_{0})\overleftrightarrow{\mathbf{M}}_{f}(\boldsymbol{\hat{R}}%
_{0})\cdot \boldsymbol{\hat{E}}_{\text{cav}}(\boldsymbol{\hat{R}}_{2})]
\notag \\
&=&-\alpha  \text{Re}[E_{10}(\boldsymbol{\hat{R}}_{1})e^{-ik_{\text{%
tw}}\hat{Z}_{1}}e^{-i\omega _{\text{tw}}t}]\boldsymbol{e}_{\text{tw}%
}^{(1)} \cdot \text{Re}[e^{ik\hat{R}_{0}}\eta _{f}(D/\hat{R}_{0})%
\overleftrightarrow{\mathbf{M}}_{f}(\boldsymbol{\hat{R}}_{0})\cdot \epsilon
_{\text{cav}}\cos (k\hat{X}_{2})(\hat{a}^{\dagger }e^{i\omega _{\text{tw}}t}+%
\hat{a}e^{-i\omega _{\text{tw}}t})\boldsymbol{e}_{\text{cav}}]  \notag \\
&\approx &-\frac{1}{2}\alpha \eta _{f}\epsilon _{\text{cav}}(D/\hat{R}%
_{0})\cos (k\hat{R}_{0})E_{10}(\boldsymbol{%
\hat{R}}_{1})\cos (k\hat{X}_{2})(\hat{a}^{\dagger }e^{-ik_{\text{tw}}\hat{Z}_{1}}+\hat{a}e^{ik_{\text{tw}}\hat{Z}_{1}})\boldsymbol{e}_{\text{tw}%
}^{(1)}\cdot \overleftrightarrow{\mathbf{M}}_{f}(\boldsymbol{\hat{R}}%
_{0})\cdot \boldsymbol{e}_{\text{cav}}, \label{A6}
\end{eqnarray}%
and%
\begin{eqnarray}
\hat{H}_{\text{tw-Gcav}}^{\left( 2\right) } &=&-\alpha \bm{\mathcal{E}}_{\text{tw2}%
}(\boldsymbol{\hat{R}}_{2},t)\cdot \text{Re}[e^{ik\hat{R}_{0}}\eta _{f}(D/%
\hat{R}_{0})\overleftrightarrow{\mathbf{M}}_{f}(\boldsymbol{\hat{R}}%
_{0})\cdot \boldsymbol{\hat{E}}_{\text{cav}}(\boldsymbol{\hat{R}}_{1})]
\notag \\
&\approx &-\frac{1}{2}\alpha \eta _{f}\epsilon _{\text{cav}}(D/\hat{R}%
_{0})\cos (k\hat{R}_{0})E_{20}(\boldsymbol{%
\hat{R}}_{2})\cos (k\hat{X}_{1})(\hat{a}^{\dagger }e^{-ik_{\text{tw}}\hat{Z}_{2}}+\hat{a}e^{ik_{\text{tw}}\hat{Z}_{2}})\boldsymbol{e}_{\text{tw}%
}^{(2)}\cdot \overleftrightarrow{\mathbf{M}}_{f}(\boldsymbol{\hat{R}}%
_{0})\cdot \boldsymbol{e}_{\text{cav}}. \label{A7}
\end{eqnarray}
Combining Eqs.~(\ref{A6}) and~{(\ref{A7})}, we can obtain the form of $\hat{H}_{\text{tw-Gcav}}$, as shown in Eq.~(\ref{HTGC1}). Then we get the approximate Hamiltonian by expanding the fields near the foci of the tweezers,
\begin{eqnarray}
\hat{H}_{\text{tw-Gcav}}&\approx &-\frac{1}{2}\alpha\eta _{f}\epsilon _{\text{cav}}\cos (kD)\varsigma( \hat{a}^{\dagger }+\hat{a}) \notag\\
&&+\frac{1}{2}\alpha \eta _{f}\epsilon _{\text{cav}}\left\{
\left[ k\sin \left( kD\right) +\cos \left( kD\right) /D\right]\varsigma
+\epsilon _{\text{tw}}^{(2)}k\cos (kD)\sin (kx_{10})%
\right\}(\hat{a}^{\dagger }+\hat{a})\hat{x}_{1} \notag\\
&& -\frac{1}{2}\alpha \eta _{f}\epsilon _{\text{cav}}\left\{
\left[ k\sin \left( kD\right) +\cos \left( kD\right) /D\right]\varsigma
-\epsilon _{\text{tw}}^{(1)}k\cos (kD)\sin (kx_{20})%
\right\}(\hat{a}^{\dagger }+\hat{a})\hat{x}_{2} \notag\\
&&-\frac{i}{2}\alpha \eta _{f}\epsilon _{\text{cav}}\epsilon _{\text{tw}%
}^{(1)}\cos (kD)k_{\text{tw}}\cos (kx_{20})(\hat{a}-\hat{a}^{\dagger })\hat{z}_{1}-\frac{i}{2}\alpha \eta _{f}\epsilon _{\text{cav}}\epsilon _{\text{tw}%
}^{(2)}\cos (kD)k_{\text{tw}}\cos (kx_{10})(\hat{a}-\hat{a}^{\dagger })\hat{z}_{2} \notag\\
&&+\mathcal{O}(\hat{Q}^{2}_{j}),
\end{eqnarray}
with $\varsigma=\epsilon _{\text{tw}}^{(1)}\cos \left( kx_{20}\right) +\epsilon _{\text{tw}}^{(2)}\cos \left( kx_{10}\right)$.
We keep the terms up to the first order of the center-of-mass displacements and obtain Eq.~(\ref{19}).

Similarly, using the expressions
 of the two electric fields, we can further express Eq.~(\ref{Hcgt}) as
\begin{eqnarray}
\hat{H}_{\text{cav-Gtw}}^{\left( 1\right) } &=&-\alpha \boldsymbol{\hat{E}}_{\text{%
cav}}(\boldsymbol{\hat{R}}_{1})\cdot\text{Re}[ e^{ik_{\text{tw}}\hat{R}%
_{0}}\eta _{f_{\text{tw}}}(D/\hat{R}_{0})\overleftrightarrow{\mathbf{M}}_{f_{%
\text{tw}}}(\boldsymbol{\hat{R}}_{0})\cdot\boldsymbol{E}_{\text{tw2}}(\boldsymbol{%
\hat{R}}_{2},t)]  \notag \\
&=&-\alpha \epsilon _{\text{cav}}\cos (k\hat{X}_{1})(\hat{a}^{\dagger
}e^{i\omega _{\text{tw}}t}+\hat{a}e^{-i\omega _{\text{tw}}t})\boldsymbol{e}_{%
\text{cav}}\cdot \text{Re}[e^{ik_{\text{tw}}\hat{R}_{0}}\eta _{f_{\text{tw}%
}}(D/\hat{R}_{0})\overleftrightarrow{\mathbf{M}}_{f_{\text{tw}}}(\boldsymbol{%
\hat{R}}_{0})\cdot E_{20}(\boldsymbol{\hat{R}}_{2})e^{-ik_{\text{tw}}\hat{Z}%
_{2}}e^{-i\omega _{\text{tw}}t}\boldsymbol{e}_{\text{tw}}^{(2)}] \notag \\
&\approx &-\frac{1}{2}\alpha \eta _{f_{\text{tw}}}\epsilon _{\text{cav}}(D/%
\hat{R}_{0})\cos(k\hat{X}_{1})E_{20}(\boldsymbol{\hat{R}}_{2})[\hat{a}^{\dagger }e^{ik_{\text{tw}}(\hat{R}_{0}-\hat{Z}_{2})}-\hat{a}e^{-ik_{\text{tw}}(\hat{R}_{0}-\hat{Z}_{2})}]\boldsymbol{e%
}_{\text{cav}}\cdot \overleftrightarrow{\mathbf{M}}_{f_{\text{tw}}}(%
\boldsymbol{\hat{R}}_{0})\cdot \boldsymbol{e}_{\text{tw}}^{(2)} , \label{A9}
\end{eqnarray}%
and%
\begin{eqnarray}
\hat{H}^{(2)}_{\text{cav-Gtw}}&=&-\alpha \boldsymbol{\hat{E}}_{\text{cav}}(\boldsymbol{\hat{R}}_{2})\cdot\text{Re%
}[ e^{ik_{\text{tw}}\hat{R}_{0}}\eta _{f_{\text{tw}}}(D/\hat{R}%
_{0})\overleftrightarrow{\mathbf{M}}_{f_{\text{tw}}}(\boldsymbol{\hat{R%
}}_{0})\cdot \boldsymbol{E}_{\text{tw1}}(\boldsymbol{\hat{R}}_{1},t)]
\notag \\
&\approx &-\frac{1}{2}\alpha\eta _{f_{\text{tw}}}\epsilon _{\text{cav}}(D/%
\hat{R}_{0})\cos(k\hat{X}_{2})E_{10}(\boldsymbol{\hat{R}}_{1})[\hat{a}^{\dagger }e^{ik_{\text{tw}}(\hat{R}_{0}-\hat{Z}_{1})}+\hat{a}e^{-ik_{\text{tw}}(\hat{R}_{0}-\hat{Z}_{1})}]\boldsymbol{e%
}_{\text{cav}}\cdot \overleftrightarrow{\mathbf{M}}_{f_{\text{tw}}}(%
\boldsymbol{\hat{R}}_{0})\cdot \boldsymbol{e}_{\text{tw}}^{(1)}. \label{A10}
\end{eqnarray}%
From Eqs.~(\ref{A9})  and~(\ref{A10}), we derive the expression of $\hat{H}_{\text{cav-Gtw}}$ as presented in Eq.~(\ref{HCGT1}). By expanding the electric fields close to the foci of the tweezers, the Hamiltonian becomes
\begin{eqnarray}
\hat{H}_{\text{cav-Gtw}}&\approx & -\frac{1}{2}\alpha \eta _{f_{\text{tw}}}\epsilon _{\text{cav}}
\varsigma(\hat{a}e^{-ik_{\text{tw}}D}+\hat{a}^{\dagger }e^{ik_{\text{tw}}D}) \notag \\
&&+\frac{1}{2}\alpha \eta _{f_{\text{tw}}}\epsilon _{\text{cav}}\left\{[(D^{-1}+i k_{\text{tw}})\varsigma+\epsilon _{\text{tw}}^{(2)}k\sin (kx_{10})%
]\hat{a}e^{-ik_{\text{tw}}D}+[(D^{-1}-i k_{\text{tw}})\varsigma+\epsilon _{\text{tw}}^{(2)}k\sin (kx_{10})]\hat{a}^{\dagger}e^{ik_{\text{tw}}D}\right\}\hat{x}_{1}\notag \\
&&+\frac{1}{2}\alpha \eta _{f_{\text{tw}}}\epsilon _{\text{cav}}\left\{[(D^{-1}+i k_{\text{tw}})\varsigma+\epsilon _{\text{tw}}^{(1)}k\sin (kx_{20})]\hat{a}e^{-ik_{\text{tw}}D}
+[(D^{-1}-i k_{\text{tw}})\varsigma+\epsilon _{\text{tw}}^{(1)}k\sin (kx_{20})]\hat{a}^{\dagger}e^{ik_{\text{tw}}D}\right\}\hat{x}_{2} \notag \\
&&-\frac{i}{2}\alpha \eta _{f_{\text{tw}}}\epsilon _{\text{cav}}\epsilon _{%
\text{tw}}^{(1)}k_{\text{tw}}\cos( kx_{20})(\hat{a}e^{-ik_{\text{tw}}D}-\hat{a}^{\dagger }e^{ik_{\text{tw}}D})\hat{z}_{1}
-\frac{i}{2}\alpha \eta _{f_{\text{tw}}}\epsilon _{\text{cav}}\epsilon _{%
\text{tw}}^{(2)}k_{\text{tw}}\cos( kx_{10})(\hat{a}e^{-ik_{\text{tw}}D}-\hat{a}^{\dagger }e^{ik_{\text{tw}}D})\hat{z}_{2} \notag \\
&&+\mathcal{O}(\hat{Q}^{2}_{j}). \label{A11}
\end{eqnarray}
Discarding the last term in Eq.~(\ref{A11}), we reach the Hamiltonian given by Eq.~(\ref{HCGT}).

\end{widetext}


\begin{thebibliography}{99}
\bibitem{Aspelmeyer2014} M. Aspelmeyer, T. J. Kippenberg, and F. Marquardt, Cavity optomechanics, Rev. Mod. Phys. \textbf{86}, 1391 (2014).

\bibitem{MAPR2014} M. Metcalfe, Applications of Cavity Optomechanics, Appl. Phys. Rev. \textbf{1}, 031105 (2014).

%%levitated system
\bibitem{JRPP2020} J. Millen, T. S. Monteiro, R. Pettit, and A. N. Vamivakas, Optomechanics with levitated particles, Rep. Prog. Phys. \textbf{83}, 026401 (2020).

\bibitem{CSC2021} C. Gonzalez-Ballestero, M. Aspelmeyer, L. Novotny, R. Quidant, and O. Romero-Isart, Levitodynamics: Levitation and control of microscopic objects in vacuum, Science \textbf{374}, eabg3027 (2021).

\bibitem{Gaxv2307} G. Winstone, M. Bhattacharya, A. A. Geraci, T. Li, P. J. Pauzauskie, and N. Vamivakas, Levitated optomechanics: A tutorial and perspective, arXiv: 2307.11858.

%%Found
\bibitem{Ask1} A. Ashkin, Acceleration and Trapping of Particles by Radiation Pressure, Phys. Rev. Lett. \textbf{24}, 156 (1970).

\bibitem{Ask2} A. Ashkin and J. Dziedzic, Optical Levitation by Radiation Pressure, Appl. Phys. Lett. \textbf{19}, 283 (1971).

\bibitem{Ask3} A. Ashkin, J. M. Dziedzic, J. E. Bjorkholm, and S. Chu, Observation of a single-beam gradient force optical trap for dielectric particles, Opt. Lett. \textbf{11}, 288 (1986).

%%atom
\bibitem{ARMP2009}  W. D. Phillips, Nobel lecture: Laser cooling and trapping of neutral atoms, Rev. Mod. Phys. \textbf{70}, 721 (1998).

%%advance
\bibitem{TN2020} T. Delord, P. Huillery, L. Nicolas, and G. H\'{e}tet, Spin-cooling of the motion of a trapped diamond, Nature (London) \textbf{580}, 56 (2020).

\bibitem{TS2010} T. Li, S. Kheifets, D. Medellin, and M. G. Raizen, Measurement of the instantaneous velocity of a Brownian particle, Science \textbf{328}, 1673 (2010).

\bibitem{US2020} U. Deli\'{c}, M. Reisenbauer, K. Dare, D. Grass, V. Vuleti\'{c}, N. Kiesel, and M. Aspelmeyer, Cooling of a levitated nano-particle to the motional quantum ground state, Science \textbf{367}, 892 (2020).

\bibitem{LN2021} L. Magrini, P. Rosenzweig, C. Bach, A. Deutschmann-Olek, S. G. Hofer, S. Hong, N. Kiesel, A. Kugi, and M. Aspelmeyer, Real-time optimal quantum control of mechanical motion at room temperature, Nature (London) \textbf{595}, 373 (2021).

\bibitem{FN2021}  F. Tebbenjohanns, M. L. Mattana, M. Rossi, M. Frimmer, and L. Novotny, Quantum control of a nanoparticle optically levitated in cryogenic free space, Nature (London) \textbf{595}, 378 (2021).

%%acceleration measurement
\bibitem{FPRA2017} F. Monteiro, S. Ghosh, A. G. Fine, and D. C. Moore, Optical levitation of 10-ng spheres with nano-$g$ acceleration sensitivity, Phys. Rev. A \textbf{96}, 063841 (2017).

\bibitem{APRA2018} A. D. Rider, C. P. Blakemore, G. Gratta, and D. C. Moore, Single-beam dielectric-microsphere trapping with optical heterodyne detection, Phys. Rev. A \textbf{97}, 013842 (2018).

%%force sensing measurement
%\bibitem{TS2010} T. Li, S. Kheifets, D. Medellin, M. G. Raizen, Measurement of the instantaneous velocity of a Brownian particle. Science \textbf{328}, 1673 (2010).

%%mass measurement
\bibitem{YPRL2020} Y. Zheng, L.-M. Zhou, Y. Dong, C.-W. Qiu, X.-D. Chen, G.-C. Guo, and F.-W. Sun, Robust Optical-Levitation-Based Metrology of Nanoparticle's Position and Mass, Phys. Rev. Lett. \textbf{124}, 223603 (2020).

%%gyroscope
\bibitem{RPRL2018}  R. Reimann, M. Doderer, E. Hebestreit, R. Diehl, M. Frimmer, D. Windey, F. Tebbenjohanns, and L. Novotny, GHz Rotation of an Optically Trapped Nanoparticle in Vacuum, Phys. Rev. Lett. \textbf{121}, 033602 (2018).

\bibitem{JPRL2018}  J. Ahn, Z. Xu, J. Bang, Y.-H. Deng, T. M. Hoang, Q. Han, R.-M. Ma, and T. Li, Optically Levitated Nanodumbbell Torsion Balance and GHz Nanomechanical Rotor, Phys. Rev. Lett. \textbf{121}, 033603 (2018).

%%quantum phenomena
\bibitem{PNAS2010} D. E. Chang, C. A. Regal, S. B. Papp, D. J. Wilson, J. Ye, O. Painter, H. J. Kimble, and P. Zoller, Cavity opto-mechanics using an optically levitated nanosphere, Proc. Natl. Acad. Sci. U.S.A. \textbf{107}, 1005 (2010).

\bibitem{NJP2010} O. Romero-Isart, M. L. Juan, R. Quidant, and J. I. Cirac, Toward quantum superposition of living organisms, New J. Phys. \textbf{12}, 033015 (2010).

\bibitem{TLNP2011} T. Li, S. Kheifets, and M. G. Raizen, Millikelvin cooling of an optically trapped microsphere in vacuum, Nat. Phys. \textbf{7}, 527 (2011).

\bibitem{APRL2010} A. A. Geraci, S. B. Papp, and J. Kitching, Short-Range Force Detection Using Optically Cooled Levitated Microspheres, Phys. Rev. Lett. \textbf{105}, 101101 (2010).

\bibitem{JPRL2012} J. Gieseler, B. Deutsch, R. Quidant, and L. Novotny, Subkelvin Parametric Feedback Cooling of a Laser-Trapped Nanoparticle, Phys. Rev. Lett. \textbf{109}, 103603 (2012).

\bibitem{NPNAS2013} N. Kiesel, F. Blaser, U. Deli\'{c}, D. Grass, R. Kaltenbaek, and M. Aspelmeyer, Cavity cooling of an optically levitated submicron particle, Proc. Natl. Acad. Sci. U.S.A. \textbf{110}, 14180 (2013).

\bibitem{JNN2014} J. Millen, T. Deesuwan, P. Barker, and J. Anders, Nanoscale temperature measurements using non-equilibrium Brownian dynamics of a levitated nanosphere, Nat. Nanotechnol. \textbf{9}, 425 (2014).

\bibitem{DQST2021} D. C. Moore and A. A. Geraci, Searching for new physics using optically levitated sensors, Quantum Sci. Technol. \textbf{6}, 014008 (2021).

\bibitem{TNP2023} T. F. Kuang, R. Huang, W. Xiong, Y. L. Zuo, X. Han, F. Nori, C.-W. Qiu, H. Luo, H. Jing, and G. Z. Xiao, Nonlinear multi-frequency phonon lasers with active levitated optomechanics, Nat. Phys. \textbf{19}, 414 (2023).


%%Development in vacuum
%%\bibitem{NJP2010} O. Romero-Isart, M. Juan, R. Quidant, and J. Cirac, Toward quantum superposition of living organisms, New J. Phys. \textbf{12}, 033015 (2010).
%%\bibitem{PNAS2010} D. E. Chang, C. A. Regal, S. B. Papp, D. J. Wilson, J. Y e, O. Painter, H. J. Kimble, and P. Zoller, Cavity opto-mechanics using an optically levitated nanosphere, Proc. Natl. Acad. Sci. U.S.A. \textbf{107}, 1005 (2010).
%%\bibitem{TLNP2011} T. Li, S. Kheifets, and M. G. Raizen, Millikelvin cooling of an optically trapped microsphere in vacuum, Nat. Phys. \textbf{7}, 527 (2011).
\bibitem{Uaxv1902} U. Deli\'{c}, D. Grass, M. Reisenbauer, T. Damm, M. Weitz, N. Kiesel, and M. Aspelmeyer, Levitated cavity optomechanics in high vacuum, Quantum Sci. Technol. \textbf{5}, 025006 (2020).

%%Cooling
\bibitem{IPRL2007} I. Wilson-Rae, N. Nooshi, W. Zwerger, and T. J. Kippenberg, Theory of Ground State Cooling of a Mechanical Oscillator Using Dynamical Backaction, Phys. Rev. Lett. \textbf{99}, 093901 (2007).
\bibitem{FPRL2007} F. Marquardt, J. P. Chen, A. A. Clerk, and S. M. Girvin, Quantum Theory of Cavity-Assisted Sideband Cooling of Mechanical Motion, Phys. Rev. Lett. \textbf{99}, 093902 (2007).

\bibitem{JN2011} J. Chan, T. P. M. Alegre, A. H. Safavi-Naeini, J. T. Hill, A. Krause, S. Gr\"{o}blacher, M. Aspelmeyer, and O. Painter, Laser cooling of a nanomechanical oscillator into its quantum ground state, Nature (London) \textbf{478}, 89 (2011).

\bibitem{CPRL2019} C. Sommer and C. Genes, Partial Optomechanical Refrigeration via Multimode Cold-Damping Feedback, Phys. Rev. Lett. \textbf{123}, 203605 (2019).

\bibitem{LiuPRA2022} Y.-H. Liu, X.-L. Yin, J.-F. Huang, and J.-Q. Liao, Accelerated ground-state cooling of an optomechanical resonator via shortcuts to adiabaticity, Phys. Rev. A \textbf{105}, 023504 (2022).

%%levitated feedback cooling
%%\bibitem{APRL2010} A. A. Geraci, S. B. Papp, and J. Kitching, Short-Range Force Detection Using Optically Cooled Levitated Microspheres, Phys. Rev. Lett. \textbf{105}, 101101 (2010).
%%\bibitem{TNP2011} T. Li, S. Kheifets, and M. G. Raizen, Millikelvin cooling of an optically trapped microsphere in vacuum, Nat. Phys. \textbf{7}, 527 (2011).
%%\bibitem{JPRL2012} J. Gieseler, B. Deutsch, R. Quidant, and L. Novotny, Subkelvin Parametric Feedback Cooling of a Laser-Trapped Nanoparticle, Phys. Rev. Lett. \textbf{109}, 103603 (2012).
\bibitem{RPRA2018} R. Diehl, E. Hebestreit, R. Reimann, F. Tebbenjohanns, M. Frimmer, and L. Novotny, Optical levitation and feedback cooling of a nanoparticle at subwavelength distances from a membrane, Phys. Rev. A \textbf{98}, 013851 (2018).

\bibitem{GPRL2019} G. P. Conangla, F. Ricci, M. T. Cuairan, A. W. Schell, N. Meyer, and R. Quidant, Optimal Feedback Cooling of a Charged Levitated Nanoparticle with Adaptive Control, Phys. Rev. Lett. \textbf{122}, 223602 (2019).

\bibitem{FPRL2019} F. Tebbenjohanns, M. Frimmer, A. Militaru, V. Jain, and L. Novotny, Cold Damping of an Optically Levitated Nanoparticle to Microkelvin Temperatures, Phys. Rev. Lett. \textbf{122}, 223601 (2019).

%%levitated sideband cooling
\bibitem{PRA2010} P. F. Barker and M. N. Shneider, Cavity cooling of an optically trapped nanoparticle, Phys. Rev. A \textbf{81}, 023826 (2010).

\bibitem{PRA2011}  O. Romero-Isart, A. C. Pflanzer, M. L. Juan, R. Quidant, N. Kiesel, M. Aspelmeyer, and J. I. Cirac, Optically levitating dielectrics in the quantum regime: Theory and protocols, Phys. Rev. A \textbf{83}, 013803 (2011).

\bibitem{PNC2013} P. Asenbaum, S. Kuhn, S. Nimmrichter, U. Sezer, and M. Arndt, Cavity cooling of free silicon nanoparticles in high vacuum, Nat. Commun. \textbf{4}, 2743 (2013).

\bibitem{JPRL2015} J. Millen, P. Z. G. Fonseca, T. Mavrogordatos, T. S. Monteiro, and P. F. Barker, Cavity Cooling a Single Charged Levitated Nanosphere, Phys. Rev. Lett. \textbf{114}, 123602 (2015).

\bibitem{PPRL2016} P. Z. G. Fonseca, E. B. Aranas, J. Millen, T. S. Monteiro, and P. F. Barker, Nonlinear Dynamics and Strong Cavity Cooling of Levitated Nanoparticles, Phys. Rev. Lett. \textbf{117}, 173602 (2016).

\bibitem{APRR2022} A. Ranfagni, K. B{\o}rkje, F. Marino, and F. Marin, Two-dimensional quantum motion of a levitated nanosphere, Phys. Rev. Research \textbf{4}, 033051 (2022).

%%actively driving
%%\bibitem{NPNAS2013} N. Kiesel, F. Blaser, U. Deli\'{c}, D. Grass, R. Kaltenbaek, and M. Aspelmeyer, Proc. Natl. Acad. Sci. U.S.A. \textbf{110}, 14180 (2013).
\bibitem{NPRL1902} N. Meyer, A. de los R\'{i}os Sommer, P. Mestres, J. Gieseler, V. Jain, L. Novotny, and R. Quidant, Resolved-Sideband Cooling of a Levitated Nanoparticle in the Presence of Laser Phase Noise, Phys. Rev. Lett. \textbf{123}, 153601 (2019).
%\bibitem{Uaxv1902} U. Deli\'{c}, D. Grass, M. Reisenbauer, T. Damm, M. Weitz, N. Kiesel, and M. Aspelmeyer, Levitated cavity optomechanics in high vacuum, Quantum Sci. Technol. \textbf{5}, 025006 (2020).

%%decrease cool rate
%%\bibitem{Uaxv1902} U. Deli\'{c}, D. Grass, M. Reisenbauer, T. Damm, M. Weitz, N. Kiesel, and M. Aspelmeyer, Levitated cavity optomechanics in high vacuum, Quantum Sci. Technol. \textbf{5}, 025006 (2020).

%%Laster phase
\bibitem{PPRA2009} P. Rabl, C. Genes, K. Hammerer, and M. Aspelmeyer, Phase-noise induced limitations on cooling and coherent evolution in optomechanical systems, Phys. Rev. A \textbf{80}, 063819 (2009).

\bibitem{ANJP2012} A. M. Jayich, J. C. Sankey, K. B{\o}rkje, D. Lee, C. Yang, M. Underwood, L. Childress, A. Petrenko, S. M. Girvin, and J. G. E. Harris, Cryogenic optomechanics with a Si$_{3}$N$_{4}$ membrane and classical laser noise, New J. Phys. \textbf{14}, 115018 (2012).

\bibitem{ANJP2013} A. H. Safavi-Naeini, J. Chan, J. T. Hill, S. Gr\"{o}blacher, H. Miao, Y. Chen, M. Aspelmeyer, and O. Painter, Laser noise in cavity-optomechanical cooling and thermometry, New J. Phys. \textbf{15}, 035007 (2013).

%%coherent scattering
\bibitem{UPRL2019} U. Deli\'{c}, M. Reisenbauer, D. Grass, N. Kiesel, V. Vuletic, and M. Aspelmeyer, Cavity Cooling of a Levitated Nanosphere by Coherent Scattering, Phys. Rev. Lett. \textbf{122}, 123602 (2019).

\bibitem{DPRL2019} D. Windey, C. Gonzalez-Ballestero, P . Maurer, L. Novotny, O. Romero-Isart, and R. Reimann, Cavity-Based 3D Cooling of a Levitated Nanoparticle via Coherent Scattering, Phys. Rev. Lett. \textbf{122}, 123601 (2019).

\bibitem{CPRA2019} C. Gonzalez-Ballestero, P. Maurer, D. Windey, L. Novotny, R. Reimann, and O. Romero-Isart, Theory for cavity cooling of levitated nanoparticles via coherent scattering: Master equation approach, Phys. Rev. A \textbf{100}, 013805 (2019).
%%\bibitem{US2020} U. Deli\'{c}, M. Reisenbauer, K. Dare, D. Grass, V. Vuleti\'{c}, N. Kiesel, and M. Aspelmeyer, Cooling of a levitated nano-particle to the motional quantum ground state, Science \textbf{367}, 892 (2020).
\bibitem{JNP2023} J. Piotrowski, D. Windey, J. Vijayan, C. Gonzalez-Ballestero, A. de los R\'{i}os Sommer, N. Meyer, R. Quidant, O. Romero-Isart, R. Reimann, and L. Novotny, Simultaneous ground-state cooling of two mechanical modes of a levitated nanoparticle, Nat. Phys.  \textbf{19}, 1009 (2023).


%%atomic physics experiments
\bibitem{VPRL2000} V. Vuleti\'{c} and S. Chu, Laser Cooling of Atoms, Ions, or Molecules by Coherent Scattering, Phys. Rev. Lett. \textbf{84}, 3787 (2000).

%%Strong coupling
\bibitem{NC2021} A. de los R\'{i}os Sommer, N. Meyer, and R. Quidant, Strong optomechanical coupling at room temperature by coherent scattering, Nat. Commun. \textbf{12}, 276 (2021).

\bibitem{Kaxv2305} K. Dare, J. J. Hansen, I. Coroli, A. Johnson, M. Aspelmeyer, and U. Deli\'{c}, Linear Ultrastrong Optomechanical Interaction, arXiv:2305.16226.

%%multiparticle
\bibitem{CPRL2017} C. Marletto and V. Vedral, Gravitationally Induced Entanglement between Two Massive Particles is Sufficient Evidence of Quantum Effects in Gravity, Phys. Rev. Lett. \textbf{119}, 240402 (2017).

\bibitem{YO2018} Y. Arita, E. M. Wright, and K. Dholakia, Optical Binding of two cooled micro-gyroscopes levitated in vacuum, Optica \textbf{5}, 910 (2018).

\bibitem{HPRA2020} H. Rudolph, K. Hornberger, and B. A. Stickler, Entangling levitated nanoparticles by coherent scattering, Phys. Rev. A \textbf{101}, 011804(R) (2020).

\bibitem{ANJP2020} A. K. Chauhan, O. \v{C}ernot\'{i}k, and R. Filip, Stationary Gaussian entanglement between levitated nanoparticles, New J. Phys. \textbf{22}, 123021 (2020).

\bibitem{VO2021} V. Svak, J. Flaj\v{s}manov\'{a}, L. Chv\'{a}tal, M. \v{S}iler, A. Jon\'{a}\v{s}, J. Je\v{z}ek, S. H. Simpson, P. Zem\'{a}nek, and O. Brzobohat\'{y}, Stochastic dynamics of optically bound matter levitated in vacuum, Optica \textbf{8}, 220 (2021).

\bibitem{YO2022} Y. Arita, G. D. Bruce, E. M. Wright, S. H. Simpson, P. Zem\'{a}nek, and K. Dholakia, All-optical sub-Kelvin sympathetic cooling of a levitated microsphere in vacuum, Optica \textbf{9}, 1000 (2022).

\bibitem{HPRL2022} H. Rudolph, U. Deli\'{c}, M. Aspelmeyer, K. Hornberger, and B. A. Stickler, Force-Gradient Sensing and Entanglement via Feedback Cooling of Interacting Nanoparticles, Phys. Rev. Lett. \textbf{129}, 193602 (2022).

\bibitem{TPRR2023} T. W. Penny, A. Pontin, and P. F. Barker, Sympathetic cooling and squeezing of two colevitated nanoparticles, Phys. Rev. Research \textbf{5}, 013070 (2023).

\bibitem{Harx2023} H. Rudolph, U. Deli\'{c}, K. Hornberger, and B. A. Stickler, Quantum theory of non-hermitian optical binding between nanoparticles, arXiv:2306.11893.

\bibitem{JVarx2023} J. Vijayan, J. Piotrowski, C. Gonzalez-Ballestero, K. Weber, O. Romero-Isart, and L. Novotny, Cavity-mediated long-range interactions in levitated optomechanics, Nat. Phys. (2024). doi:10.1038/s41567-024-02405-3.

\bibitem{JNN2023} J. Vijayan, Z. Zhang, J. Piotrowski, D. Windey, F. van der Laan, M. Frimmer, and L. Novotny, Scalable all-optical cold damping of levitated nanoparticles, Nat. Nanotechnol. \textbf{18}, 49 (2023).

\bibitem{Marx2023} M. Reisenbauer, H. Rudolph, L. Egyed, K. Hornberger, A. V. Zasedatelev, M. Abuzarli, B. A. Stickler, and U. Deli\'{c}, Non-Hermitian dynamics and nonreciprocity of optically coupled nanoparticles, arXiv:2310.02610.

\bibitem{Varx2023} V. Li\v{s}ka, T. Zem\'{a}nkov\'{a}, P. J\'{a}kl, M. \v{S}iler, S. H. Simpson, P. Zem\'{a}nek, and O. Brzobohat\'{y}, Observations of a PT phase transition and collective limit cycle oscillations in non-reciprocally coupled optomechanical oscillators levitated in vacuum, arXiv:2310.03701.

\bibitem{CarX2023} C. Jakubec, P. Solano, U. Deli\'{c}, and K. Sinha, Fluctuation-induced forces on nanospheres in external fields, Phys. Rev. A \textbf{109}, 052807 (2024).


%%array
\bibitem{SA2020} S. Liu, Z.-q. Yin, and T. Li, Prethermalization and nonreciprocal phonon transport in a levitated optomechanical array, Adv. Quantum Technol. \textbf{3}, 1900099 (2020).

\bibitem{ANPJ2022} A. K. Chauhan, O. \v{C}ernot\'{i}k, and R. Filip, Tuneable Gaussian entanglement in levitated nanoparticle arrays, npj Quantum Inf. \textbf{8}, 151 (2022).

\bibitem{JPR2023} J. Yan, X. Yu, Z. V. Han, T. Li, and J. Zhang, On-demand assembly of optically-levitated nanoparticle arrays in vacuum, Photon. Res. \textbf{11}, 600 (2023).

\bibitem{YSC2023}  Y. Bao, S. S. Yu, L. Anderegg, E. Chae, W. Ketterle, K.-K. Ni, and J. M. Doyle, Dipolar spin-exchange and entanglement between molecules in an optical tweezer array, Science \textbf{382}, 1138 (2023).

\bibitem{CSC2023} C. M. Holland, Y. Lu, and L. W. Cheuk, On-demand entanglement of molecules in a reconfigurable optical tweezer array, Science \textbf{382}, 1143 (2023).

%%Optical binding
\bibitem{MPRL1989} M. M. Burns, J.-M. Fournier, and J. A. Golovchenko, Optical binding, Phys. Rev. Lett. \textbf{63}, 1233 (1989).

\bibitem{VARB2006} V. Kar\'{a}sek, K. Dholakia, and P. Zem\'{a}nek, Analysis of optical binding in one dimension, Appl. Phys. B \textbf{84}, 149 (2006).

\bibitem{KRMP2010} K. Dholakia and P. Zem\'{a}nek, Colloquium: Gripped by light: Optical binding, Rev. Mod. Phys. \textbf{82}, 1767 (2010).

\bibitem{JS2022} J. Rieser, M. A. Ciampini, H. Rudolph, N. Kiesel, K. Hornberger, B. A. Stickler, M. Aspelmeyer, and U. Deli\'{c}, Tunable light-induced dipole-dipole interaction between optically levitated nanoparticles, Science \textbf{377}, 987 (2022).

%%interaction Hamiltonian
%%\bibitem{UPRL2019} U. Deli\'{c}, M. Reisenbauer, D. Grass, N. Kiesel, V. Vuleti 锟斤拷c, and M. Aspelmeyer, Cavity Cooling of a Levitated Nanosphere by Coherent Scattering, Phys. Rev. Lett. \textbf{122}, 123602 (2019).
%%\bibitem{CPRA2019} C. Gonzalez-Ballestero, P. Maurer, D. Windey, L. Novotny, R. Reimann, and O. Romero-Isart, Theory for cavity cooling of levitated nanoparticles via coherent scattering: Master equation approach, Phys. Rev. A \textbf{100}, 013805 (2019).

%%Green function
%%\bibitem{KRMP2010} K. Dholakia and P. Zem\'{a}nek, Colloquium: Gripped by light: Optical binding, Rev. Mod. Phys. \textbf{82}, 1767 (2010).
\bibitem{Lbook2012} L. Novotny and B. Hecht, \emph{Principles of Nano-optics} (Cambridge University Press, Cambridge, Eangland, 2012).
\bibitem{MPRA2018} M. A. Abbassi and K. Mehrany, Inclusion of the backaction term in the total optical force exerted upon rayleigh particles in nonresonant structures, Phys. Rev. A \textbf{98}, 013806 (2018).

%%far-field
\bibitem{Daxv2203} D. De Bernardis, G. Rastelli, I. Carusotto, and V. Scarani, Optical-force-mediated coupling between levitated nanospheres can go ultrastrong, arXiv:2203.10126.

%%correlation functions
\bibitem{Cbook2000} C. W. Gardiner and P. Zoller, \emph{Quantum Noise} (Springer, Berlin, 2000).

%%the Lyapunov equation
\bibitem{DPRL2007} D. Vitali, S. Gigan, A. Ferreira, H. R. Bohm, P. Tombesi, A. Guerreiro, V. Vedral, A. Zeilinger, and M. Aspelmeyer, Optomechanical Entanglement between a Movable Mirror and a Cavity Field, Phys. Rev. Lett. \textbf{98}, 030405 (2007).
%%the Routh-Hurwitz criterion
\bibitem{Ibook2014} I. S. Gradshteyn and I. M. Ryzhik, \emph{Table of Integrals, Series, and Products} (Academic, New York, 2014).

%%dark mode
%%\bibitem{JNP2023} J. Piotrowski, D. Windey, J. Vijayan, C. Gonzalez-Ballestero, A. de los R\'{i}os Sommer, N. Meyer, R. Quidant, O. Romero-Isart, R. Reimann, and L. Novotny, Simultaneous ground-state cooling of two mechanical modes of a levitated nanoparticle, Nat. Phys.  \textbf{19}, 1009 (2023).
\bibitem{CNJP2008} C. Genes, D. Vitali, and P. Tombesi, Simultaneous cooling and entanglement of mechanical modes of a micromirror in an optical cavity, New J. Phys. \textbf{10}, 095009 (2008).
\bibitem{WangPRL2012} Y.-D. Wang and A. A. Clerk, Using Interference for High Fidelity Quantum State Transfer in Optomechanics, Phys. Rev. Lett. \textbf{108}, 153603 (2012).
\bibitem{TianPRL2012} L. Tian, Adiabatic State Conversion and Pulse Transmission in Optomechanical Systems, Phys. Rev. Lett. \textbf{108}, 153604 (2012).


\bibitem{DPRA2020} D.-G. Lai, J.-F. Huang, X.-L. Yin, B.-P. Hou, W. Li, D. Vitali, F. Nori, and J.-Q. Liao, Nonreciprocal ground-state cooling of multiple mechanical resonators, Phys. Rev. A \textbf{102}, 011502(R) (2020).

\bibitem{DPRL2022} D.-G. Lai, J.-Q. Liao, A. Miranowicz, and F. Nori, Noise-Tolerant Optomechanical Entanglement Via Synthetic Magnetism, Phys. Rev. Lett. \textbf{129}, 063602 (2022).

\bibitem{JPRA2022a} J. Huang, D.-G. Lai, C. Liu, J.-F. Huang, F. Nori, and J.-Q. Liao, Multimode optomechanical cooling via general dark-mode control, Phys. Rev. A \textbf{106}, 013526 (2022).

\bibitem{Jarx2023} J. Huang, C. Liu, X.-W. Xu, and J.-Q. Liao, Dark-Mode Theorems for Quantum Networks, arXiv:2312.06274.
%\bibitem{JPRA2022b} J. Huang, D.-G. Lai, and J.-Q. Liao, Thermal-noise-resistant optomechanical entanglement via general dark-mode control, Phys. Rev. A \textbf{106}, 063506 (2022).

%%Q factors 10^{8}
\bibitem{JNP2013} J. Gieseler, L. Novotny, and R. Quidant, Thermal nonlinearities in a nanomechanical oscillator, Nat. Phys. \textbf{9}, 806 (2013).
%%small kappa
\bibitem{CPRA2008} C. Genes, D. Vitali, P. Tombesi, S. Gigan, and M. Aspelmeyer, Ground-state cooling of a micromechanical oscillator: Comparing cold damping and cavity-assisted cooling schemes, Phys. Rev. A \textbf{77}, 033804 (2008).
\bibitem{MPRL2018} M. Rossi, N. Kralj, S. Zippilli, R. Natali, A. Borrielli, G. Pandraud, E. Serra, G. Di Giuseppe, and D. Vitali, Normal-Mode Splitting in a Weakly Coupled Optomechanical System, Phys. Rev. Lett. \textbf{120}, 073601 (2018).
%%recoil heating
\bibitem{VPRL2016} V. Jain, J. Gieseler, C. Moritz, C. Dellago, R. Quidant, and L. Novotny, Direct Measurement of Photon Recoil from a Levitated Nanoparticle, Phys. Rev. Lett. \textbf{116}, 243601 (2016).

\end{thebibliography}
\end{document}